\documentclass[acmsmall]{acmart}
\usepackage{hyperref}
\AtBeginDocument{%
  }

\setcopyright{cc}
\setcctype{by-nc-nd}
\acmJournal{PACMHCI}
\acmYear{2026} \acmVolume{10}
\acmNumber{6}\acmArticle{CSCW186}
\acmMonth{10}\acmPrice{}
\acmDOI{10.1145/3817034}
\usepackage[normalem]{ulem}    
\renewcommand{\sout}[1]{}
\colorlet{blue}{black}
\begin{document}

\title[We are all in big trouble!]{``We are all in big trouble! *Shock Emoji'': Personal Narratives in Expressing Emotions, Opinions, and Data Regarding Climate Change in  TikTok Short Videos}


\author{Chu Zhang}
\email{zhangchu0908@outlook.com}
\orcid{0009-0004-4491-2279}
\affiliation{
\institution{Guangdong University of Technology}
\city{Guangzhou}
\country{China}}

\author{Simai Huang}
\email{huangstella0124@gmail.com}
\orcid{0009-0004-7576-0220}
\affiliation{
\institution{City University of Hong Kong}
\city{Hong Kong, SAR}
\country{China}}

\author{Shaohua Wu}
\email{ShaohuaWu26@outlook.com}
\orcid{0009-0008-6042-3351}
\affiliation{
\institution{The University of Hong Kong}
\city{Hong Kong, SAR}
\country{China}}

\author{Yihuan Chen}
\email{yihuachen6-c@my.cityu.edu.hk}
\orcid{0009-0004-2723-7398}
\affiliation{
\institution{City University of Hong Kong}
\city{Hong Kong, SAR}
\country{China}}

\author{RAY LC}
\authornote{Correspondences can be addressed to ray.lc@cityu.edu.hk.}
\email{ray.lc@cityu.edu.hk}
\orcid{0000-0001-7310-8790}
\affiliation{
\institution{City University of Hong Kong\\Studio for Narrative Spaces}
\city{Hong Kong, SAR}
\country{China}}

\renewcommand{\shortauthors}{Zhang, et al.}


\begin{abstract}
Climate change is a source of anxiety about the future. Understanding how people express themselves about climate change enables us to address such concerns. To study climate change expression on social media, we analyzed 200 TikTok videos tagged with \#climatechange, identifying four categories of content: expression-feelings, views-appeals, news-information, and trend-hijacking. We found that creators use humor to package sharp critiques, avoiding direct confrontation. They replace complex discussions with life stories, such as adopting a vegetarian lifestyle or deleting emails. They borrow from news media to present fragmented information as scientific interpretations, creating a perception of scientific credibility, \textcolor{blue}{balancing scientific accuracy with emotionality}. Analysis of viewer responses showed they engaged empathetically, reshaping interpretations of videos. \textcolor{blue}{These interactions risk reinforcing existing views but help build community on TikTok, which lacks community structure.} This study reveals how creators may retell news on science using personal narratives, \textcolor{blue}{highlighting how short-form videos enable climate communication}.
\end{abstract}



\received{May 13, 2025}
\received[revised]{January 13, 2026}
\received[accepted]{April 9, 2026}

\maketitle

\section{Introduction}\label{sec:Introduction}
In climate communication, awareness is easy, but action is hard. Translating awareness of the climate crisis into effective policy and public action is difficult, because people often do not envision the long-term consequences of climate destruction~\cite{singh2017perceived}. Research highlights challenges in communicating climate change to the public, revealing a gap in understanding \textcolor{blue}{of} how to effectively raise awareness and emotionally engage the audience \textcolor{blue}{in addressing} the climate crisis~\cite{moser2016reflections, wang2018emotions}. Effective climate communication requires collaboration and coordination among all parties to ensure that diverse perspectives and concerns are addressed, and that feasible solutions are widely understood and adopted.

Various strategies have been used to enhance public understanding of climate change through the dissemination of science, policy, and statistics~\cite{de2021transforming}. However, audience responses, along with their concerns and discussions about climate change, are often overlooked in these efforts~\cite{verlie2022learning}. Social media platforms, with their diverse features, such as popular music, special effects, and editing, offer content creators multiple ways of expression~\cite{omar, klug}, while also providing valuable insights into public discussions on climate change~\cite{Moser}.

Although climate communication is well-studied in traditional news and science media~\cite{Hautea, Yunpeng}, how content creators adopt diverse strategies to express climate issues and generate communication on social media platforms remains underexplored. With TikTok's increasingly diverse creator base, the platform uniquely blends entertainment with substantive scientific content~\cite{Schaadhardt, Markazi}, offering a rich landscape for climate communication research. Climate communication theory focuses on how diverse channels and visual-interactive strategies influence the reach and impact of climate messaging. Existing studies have shown that visual elements significantly enhance the persuasiveness and impact of climate change messages~\cite{nyhan2019roles}, while interaction and feedback help evaluate message effectiveness~\cite{shome2009psychology, moser2010communicating}. Therefore, this study, informed by the core tenets of climate communication theory, focuses on the information elements, modes of expression, and audience responses in TikTok videos related to climate change, providing a more concrete understanding of climate communication in social media contexts.

This study collected 3,086 videos under the \textcolor{blue}{\#climatechange} hashtag on TikTok from the past five years, randomly selecting 200 videos for analysis. We established the final taxonomy by integrating iterative open coding with the \textit{Climate Communication Visuals Framework (CCVF)}~\cite{ClimateVisuals}. This taxonomy was subsequently used to categorize the full sample of 200 videos. Following this, we randomly selected 25 videos from each category for qualitative deep analysis. In this process, we examined video length, media elements (such as images, text, and video clips), narrative content, and the popular interactions and comments left by audiences. Therefore, based on climate communication theory (which emphasizes communication channels, information features, communication strategies, and audience interaction), and the existing research gap concerning TikTok's climate change community, we address the following questions:

\begin{itemize}

    \item \textit{RQ1: What types of videos are shown by content creators on TikTok to express and inform about climate change?}

    \item \textit{RQ2: What strategies do we see content creators employ for expressing and informing about climate change in different types of videos on TikTok?}

    \item \textit{RQ3: How do viewers and creators engage collaboratively in the form of comments in response to TikTok climate change content?}

\end{itemize}

Through three interconnected RQs, this study also reveals the intertwined logic between the drive for \textcolor{blue}{media incentives} and the mandate for accurate scientific communication in TikTok's climate discourse.

\textcolor{blue}{Our results categorize} TikTok \textcolor{blue}{\#climatechange} videos into four types: \textit{Expression and Feelings}, \textit{Views and Appeals}, \textit{News and Information}, and \textit{Trend Hijacking}. These videos use visuals, sound, and emotional variation to enhance communication. For instance, \textcolor{blue}{content creators} use humor and emotional performances to highlight social issues, introducing sharp topics like governments or public figures in a seemingly harmless manner. Regardless of whether content creators are directly affected by climate change, they personalize the issue by incorporating their own values, bringing it closer to their audience. To manage complexity, creators try to package scientific knowledge in more vivid ways, such as through animations and expressions, thereby establishing their authority in the realm of knowledge. Moreover, creators and viewers interact through emotional replication, questioning, supplementing, extending, and counter-questioning in the comment section, jointly breaking down complex or abstract climate topics into discussable, concrete issues. The study also found that while official content attracts viewers, it suppresses interaction. Regardless of the expression strategy, these \textcolor{blue}{content creators} excel at building personal narratives. 

Whether through content creation to attract viewers or through audience interaction to enhance video dissemination, this process can increase the visibility of climate communication topics under the influence of platform algorithms. Our study shows that TikTok climate communication prioritizes issue vitality through engagement while maintaining scientific information. This phenomenon provides insights into \textcolor{blue}{climate communication} strategies and contributes to the development of better climate communication approaches.

\section{Background}\label{sec:Background}
\subsection{Information Gaps in Climate Communication}\

Understanding the public's knowledge, concerns, participation, and collective responses is a key objective of climate communication~\cite{climatechangecommunication}. The scientific community has already reached broad consensus on climate change~\cite{NASA, NOAA}, and many communication initiatives have focused on disseminating scientific facts~\cite{ipcc} to improve public literacy on the issue. Several studies have explored this goal through approaches such as organizing climate dialogues via workshops~\cite{clark2023community}, utilizing data visualization targeted at non-expert audiences~\cite{schuster2024being}, and integrating gaming and immersive media experiences~\cite{galeote2023text}.

However, a significant gap remains between public understanding and scientific consensus on climate change~\cite{moser2010communicating}. Traditional climate communication tends to focus on macro-level discussions within scientific and policy communities, often lacking relevance to the public's everyday experiences and immediacy~\cite{Markazi}. Many individuals report \textcolor{blue}{that they feel} disconnected from the impacts of climate change and believe they will remain largely unaffected~\cite{EPCC}. Although an increasing number of people recognize the growing relevance of climate change to society~\cite{degroot2021towards}, communication strategies that rely heavily on data and technical language often overlook emotional resonance and everyday experiences. This makes climate information difficult for the public to understand and internalize effectively~\cite{Schafer}, resulting in feelings of detachment and helplessness regarding climate change~\cite{EPCC}. Therefore, the way information is conveyed and how the public perceives and experiences that information play a critical role in shaping the effectiveness of climate communication and influencing public willingness to engage~\cite{wcc2010, de2021transforming}.

\textcolor{red}{\sout{Effective climate communication depends on how information is presented and delivered, as well as how the public perceives and experiences it. }}\textcolor{blue}{To better understand these gaps, it is useful to examine the theoretical foundations of climate communication.} Climate communication theory is an interdisciplinary framework that integrates perspectives from multiple fields. It focuses on how diverse communication approaches can effectively disseminate climate change information and enhance public knowledge, attitudes, and behaviors. The theory highlights four key aspects that shape communication outcomes: communication channels, visual elements, communication strategies, and audience interaction. Communication channels determine the reach of information and the speed of dissemination~\cite{mavrodieva2019role}. Visual elements in different forms can trigger emotional responses and increase persuasiveness~\cite{o2014climate, nyhan2019roles}. Communication strategies influence the level of public attention and willingness to take action~\cite{weber2010shapes}. Audience interaction is also essential, as it enhances participation and provides feedback for evaluating communication effectiveness~\cite{moser2010communicating}.

However, effective climate communication still faces two major challenges. The first is the cognitive barrier: climate-related information is often presented through data-heavy reports or abstract models, making it difficult for non-experts to comprehend and leading to confusion and disengagement~\cite{bayes2023research}. The second is the emotional barrier: climate change is frequently portrayed as a distant, large-scale, or future-oriented issue, making it difficult for individuals to connect it to their immediate personal lives, which reduces their sense of urgency and willingness to participate~\cite{EPCC, Lei}. Therefore, gaining a deeper understanding of how the public perceives and expresses their attitudes toward climate change is essential for designing more targeted and effective communication strategies.

\subsection{The Influence of Social Media on Climate Change Topics}\

Social media plays an increasingly important role in climate communication, serving as a dynamic space for both information dissemination and public discourse~\cite{Tapio}. Platforms like \textcolor{blue}{X (formerly Twitter)}~\cite{Schafer}, Facebook~\cite{Habib}, TikTok~\cite{shutsko2020user}, and Instagram~\cite{Suse}, amplify the reach of narratives and accelerate the spread of climate-related information~\cite{León, Wang2021}. By lowering participation barriers—such as liking, sharing, commenting, and template creation—social media encourages user expression and engagement~\cite{omar, klug}. Compared to traditional media, social media offers greater interactivity, where users are not only recipients of information but also creators and disseminators of content~\cite{Schafer}. Moreover, social media influencers are increasingly replacing traditional news media as a reliable source of information on highly politicized topics~\cite{Beers}. Additionally, these platforms provide access to content that may be difficult to obtain through traditional news channels~\cite{Dailey}, such as personal experiences shared by users. This interactivity makes social media a valuable environment for mobilizing public action and raising awareness about climate change~\cite{fernandez2016talking}.

Therefore, social media not only provides a platform for effective communication on climate issues but also plays a crucial role in sustaining public attention and fostering continuous discourse on climate issues~\cite{omar, klug}. Among younger generations, social media has become a crucial channel for expressing concerns and organizing collective action. It presents, to some extent, public emotions and opinions, offering valuable references for understanding diverse perspectives on climate change. Although the effectiveness of social media in driving behavioral and attitudinal change remains subject to further investigation—particularly given that algorithm-driven personalized recommendations may lead to \textcolor{blue}{``information bubbles,''\footnote{\textcolor{blue}{A constrained information environment in which a user’s exposure to diverse perspectives is restricted. This phenomenon emerges from the complex interplay between algorithmic filtering (e.g., Filter Bubbles) and individual preferences (e.g., Echo Chambers).}}} which limit users' exposure to diverse viewpoints, and reinforce existing climate beliefs—this phenomenon does not preclude social media from reflecting meaningful public opinions. Existing studies generally agree that, with its wide-reaching dissemination capacity and strong emotional mobilization effects, social media remains an essential platform for observing public attention trends, understanding diverse perspectives on climate issues, and facilitating ongoing public discourse~\cite{Pearce}.

In summary, social media possesses significant power in spreading information and mobilizing emotions. It shows great potential in climate communication and in encouraging public debate on climate issues. Beyond these broad features, there is still a gap \textcolor{blue}{in understanding} how creators use emotional resonance, storytelling, and interactivity to address climate topics. \textcolor{red}{\sout{If this gap is ignored, opportunities to strengthen the impact of climate communication and to raise public participation may be lost. This would also hold back innovation and reduce the effectiveness of climate communication strategies.}}\textcolor{blue}{Ignoring this gap risks losing opportunities to strengthen the impact of climate communication and enhance public participation, thereby hindering innovation and reducing the effectiveness of communication strategies.}

\subsection{The Climate Change Ecosystem on TikTok}\

Among various social media platforms, TikTok has rapidly emerged as one of the most popular short video platforms globally since its launch in 2017~\cite{Statista}. Studies show that compared to platforms like YouTube and Instagram, TikTok attracts users through its visual and musical elements, enhancing the accessibility and spread of information~\cite{Zmarzlińska}. Additionally, TikTok content creation serves as a means of personal expression~\cite{Hautea} and is an essential form of social interaction. Unlike traditional social media platforms that rely on users' social networks, TikTok uses its ``For You Page'' (FYP) algorithm to recommend content based on interaction patterns. This allows climate-related topics to reach a wider audience, including those who are not environmentalists or policy experts~\cite{Yunpeng, zulli2022extending}. The recommendation mechanism, combined with the viral nature of short videos, enables climate content to break out of its original circles and achieve rapid spread and high visibility.

TikTok's short-video format and algorithm-driven content discovery also improve the platform's usability and \textcolor{blue}{user engagement}. At the same time, they provide users with creative and engaging ways to express themselves~\cite{Herrman, Simpson}. As a result, complex topics such as climate change become more accessible. As TikTok's platform incentive structure prioritizes \textcolor{blue}{user engagement} (e.g., likes, comments, and shares)~\cite{tiktok2020tiktok}, videos often utilize trending music, dancing, or highly provocative visuals to persuade viewers. Research has explored the impact of climate change content on TikTok, such as the sustainability of messages shared by eco-influencers~\cite{Huber} and the relationship between \textcolor{blue}{user engagement} and specific video themes~\cite{Basch}. TikTok's focus on entertainment, visual appeal, and rapid dissemination offers opportunities for viral environmental content~\cite{Le}. Meanwhile, typical interface designs like infinite scrolling and real-time social feedback further strengthen user emotional arousal and engagement~\cite{herrman2019tiktok}.

Through short videos, users can vividly express their views, share experiences, and participate in public discussions~\cite{Le}. Short videos hold significant potential for enhancing public awareness of climate change, especially in their ability to convey emotions and information in real-time~\cite{León}. Consequently, this communication process may transform climate change from a pure scientific topic into a form of personal positional expression on TikTok~\cite{efstratiou2024polarisation}. To thoroughly analyze TikTok's \textcolor{blue}{climate change ecosystem}, it is necessary not only to track popular hashtags but also to examine content visualization and how videos reflect users' concerns and needs. Studying the dissemination of climate-related topics on TikTok can provide deeper insights into the platform's unique climate discourse characteristics~\cite{Le}. Strengthening our understanding of the interaction between storytelling and public response will provide valuable insights for improving online climate communication strategies~\cite{Pera}. Although existing research has examined climate-related content on social media, this study takes TikTok as the case platform and applies the framework of climate communication theory \textcolor{blue}{to analyze} information elements, expressive strategies, and audience responses in climate-related videos. In doing so, it explores the role of creator expression and comment interaction in climate communication, deepens the understanding of public participation patterns, and offers practical insights for optimizing climate communication strategies.

\section{Methods}\label{sec:Methods}
\subsection{Data Collection}
We selected the most \textcolor{blue}{represented} hashtag by analyzing frequently cited terms in TikTok climate research~\cite{Ha2023, Hautea}, such as \#climatechange, \#forclimate, and \#climateaction. \textcolor{blue}{Data extraction for each hashtag revealed that} \#climatechange is the most prominent, with 489.3k videos and 8.5 billion views. Unlike \#forclimate and \#climateaction, which are often linked to specific campaigns and show pronounced temporal spikes, \#climatechange functions as a generic, continuously used label across diverse communities. Following established CSCW practices~\cite{pera2024shifting, Markazi}, we selected \#climatechange as our primary data source.

We used the TikHub API~\cite{TikHub} to crawl 4,183 climate-related TikTok videos posted over the past five years. Due to API limitations, the dataset represents a snapshot of accessible videos rather than a comprehensive or top-viewed sample. After removing duplicates and static images (0-second videos), the final sample comprised 3,086 videos ranging from 5 seconds to 10 minutes. We extracted metadata for each video, including descriptions, tags, and engagement metrics (e.g., likes, comments, shares; see Appendix \ref{sec:Appendix}).

\subsubsection{Ethical Considerations}\

According to TikTok's Community Guidelines\footnote{https://www.tiktok.com/community-guidelines/en/overview} and Privacy Policy\footnote{https://www.tiktok.com/legal/page/us/privacy-policy/en}, all publicly posted videos and comments are accessible, shareable, and commentable by other users, meaning that the TikTok videos and comments analyzed in this study are generally considered publicly available content. Users are aware of these guidelines regarding privacy and public content. Additionally, we obtained IRB approval for conducting research involving social media content.

However, we acknowledge that the public availability of data does not automatically equate to ethical use. TikTok creators may not anticipate researchers as part of their imagined audience~\cite{sarikakis2017social}, nor do they necessarily expect their content to be repurposed beyond its original context~\cite{fiesler2018participant, klassen2022isn}. To honor users' contextual privacy expectations and mitigate potential harm, we followed specific procedures to de-identify the analyzed content:

First, we avoided using verbatim quotes that could be reverse-searched to identify specific users; instead, content is reported through paraphrasing or in aggregate form~\cite{taylor2024carefully}. Second, in all visualizations, identifiable information (such as usernames and profile images) has been anonymized through blurring. Finally, regarding highly identifiable video screenshots, we contacted the original creators individually and anonymously to seek informed consent; screenshots were excluded from the paper if consent was denied.

\subsection{Data Analysis}

To answer RQ1, we first developed a taxonomy to identify the primary characteristics of video content. We then expanded our analysis to a set of 200 videos (see Section \ref{sec:4.1}). Our findings reveal that climate change content on TikTok is categorized into four distinct types: \textit{Expression and Feelings}, \textit{Views and Appeals}, \textit{News and Information}, and \textit{Trend Hijacking}, ranging from personal narratives to factual reporting.

Based on this classification, and to address RQ2, we randomly selected 25 videos from each category for in-depth analysis. Our primary focus was on analyzing categories directly related to climate change (see Section \ref{sec:4.2}). Our results indicate that creators strategically use different media elements to achieve persuasive impacts, such as using humorous animations for emotional arousal or authoritative datasets for objective reporting.

To answer RQ3, we simultaneously conducted qualitative analysis of the comment sections during the video content analysis. Specifically, we focused on the top 6 most liked comments and those involving creator-viewer interaction (see Section \ref{sec:4.3}). Our results suggest that comments act as a discursive space for negotiating accuracy and expressing stances, often echoing or challenging the video’s emotional tone.

\subsubsection{Categorizing Videos \textcolor{blue}{Tagged} with \#climatechange}\

\setlength{\textfloatsep}{5pt}
\begin{figure}[t]
    \centering
    \includegraphics[width=1\textwidth]{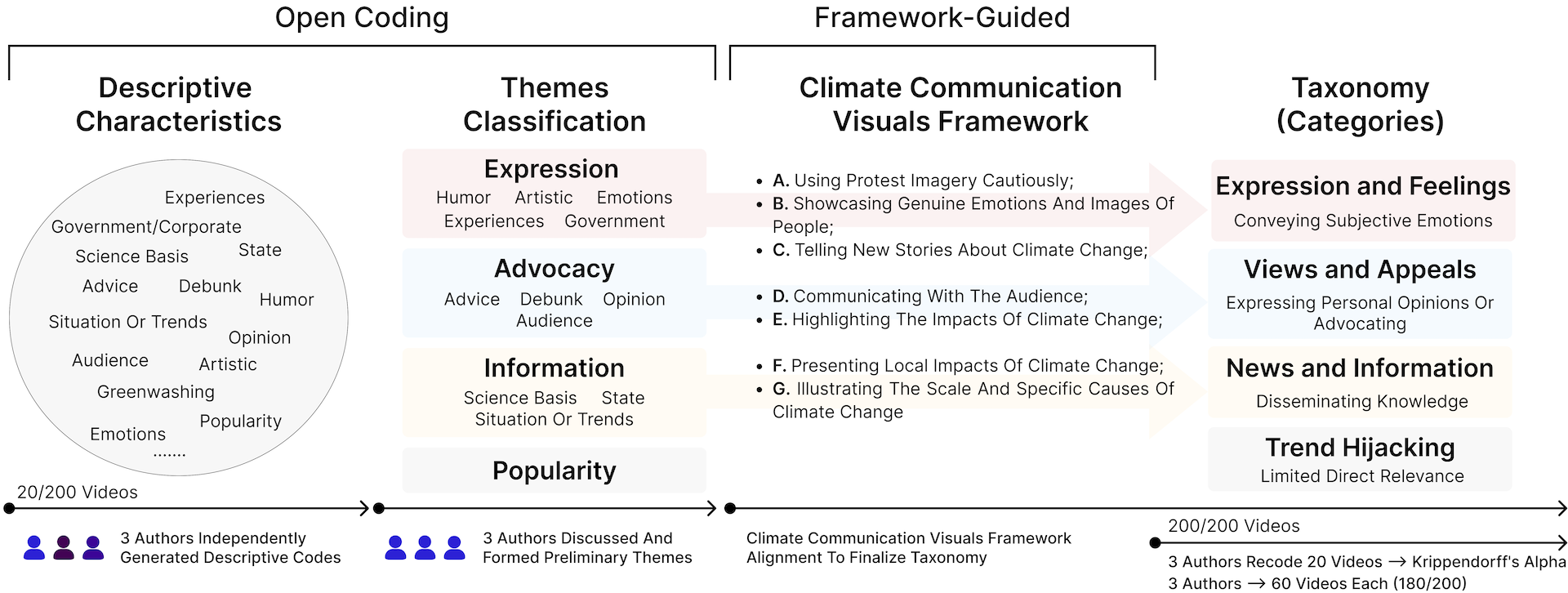}
   \caption{Steps to establish a taxonomy of climate change expression on TikTok}
    \label{fignew1}
\end{figure}

To establish the taxonomy, we used a mixed-coding approach, combining induction (data-driven) and deduction (theory-guided). Three authors first independently coded 20 videos, developing a preliminary framework through iterative discussion. For instance, we initially identified codes such as \textit{artistic, state, experience, debunk, and opinion}. The three authors then categorized these codes based on content similarity and the videos' inherent logic. For example, codes like \textit{artistic, emotion, experience, and humor} were grouped into a theme centered on personal expression and emotional presentation (\textcolor{blue}{Figure} \ref{fignew1}).

\textcolor{red}{\sout{As initial codes were too abstract for communication strategy analysis, we aligned these grouped themes with the seven key elements of the \textit{Climate Communication Visuals Framework (CCVF)}~\cite{ClimateVisuals}.}} \textcolor{blue}{As initial codes captured the richness of the videos' surface-level perceptions and intents, they lacked the aggregation of deeper strategic orientations required for a large-scale systematic analysis. Subsequently, we aligned these grouped themes with the seven key elements of the \textit{Climate Communication Visuals Framework (CCVF)} \cite{ClimateVisuals}.} These seven elements include: a. using protest imagery cautiously; b. showcasing genuine emotions and images of people; c. telling new stories about climate change; d. communicating with the audience; e. highlighting the impacts of climate change; f. presenting local impacts of climate change; and g. illustrating the scale and specific causes of climate change.

\textcolor{red}{\sout{We observed that the theme focusing on personal expression and emotional presentation was highly consistent with CCVF elements a, b, and c. This principle was used to refine and map all our codes (Figure \ref{fignew1}).}} \textcolor{blue}{We found that the initial codes and the CCVF exhibited high functional congruence in terms of communicative intent. Specifically, while initial codes such as \textit{artistic, humor, and experiences} differ in their forms of expression, they all avoid the protest imagery commonly seen in climate communication (a), achieving personalized, non-confrontational affective resonance through the establishment of genuine emotion (b) and storytelling (c). Initial codes such as \textit{debunk, opinion, and advice}, although different in their rhetorical modes, all point toward two-way communication (d) and the strengthening of climate impacts (e), aiming to persuade the viewers and convey specific stances. Initial codes such as \textit{state and science basis}, while different in their presentation, all functionally serve the construction of objective facts, achieving information gain through the presentation of local impacts (f) and scientific causes (g).} 

The CCVF served as a theoretical foundation for organizing and aggregating our exploratory findings into a classification of communication strategies (\textcolor{blue}{Figure} \ref{fignew1}). Specifically, when a video predominantly includes elements a, b or c, we classified it as \textit{Expression and Feelings}; when the video primarily embodies elements d or e, it was classified as \textit{Views and Appeals}; and when the video emphasizes elements f or g, it was classified as \textit{News and Information}. Additionally, videos that were tagged with \#climatechange but were completely unrelated to the topic of climate change were categorized as \textit{Trend Hijacking}. These four types constitute our final taxonomy.

Using this four-category taxonomy as a closed codebook, three researchers re-coded the initial 20 videos. To verify the consistency of the coding, we measured the reliability of the three coders' classification of the 20 videos using Krippendorff's Alpha. The result was 0.81, indicating a high level of agreement between the coders ($\alpha > 0.80$). After verification, the remaining 180 videos were independently coded by the three co-authors (each responsible for 60 videos) to ensure the comprehensiveness and consistency of the classification. In this process, we referenced sampling methods used in other video analysis studies~\cite{Wang2024Heritage}, and confirmed that all 200 videos could be categorized into the four types, confirming the taxonomy's exhaustiveness.

\subsubsection{Analyses of Video Content and Viewers Responses}\

To further understand how creators express climate change, we employed a comprehensive content analysis across different video types. While a larger sample could improve precision, the time-intensive nature of deep content analysis yields diminishing returns beyond a certain point. The classification distribution ratio for the four video types is 42:58:57:43. To balance depth and efficiency, we randomly selected 25 videos from each category. Quantitative research in content analysis indicates that a sample size of 25 provides sufficient depth and diversity to address our research questions while maintaining methodological rigor~\cite{creswell2017research}. Unavailable videos were replaced with new random selections from the same categories to ensure a consistent sample size. Future research could expand the study's scope within TikTok (e.g., by incorporating a broader range of hashtags) to enhance the precision of the findings.

\begin{figure}[t]
    \centering
    \includegraphics[width=1\textwidth]{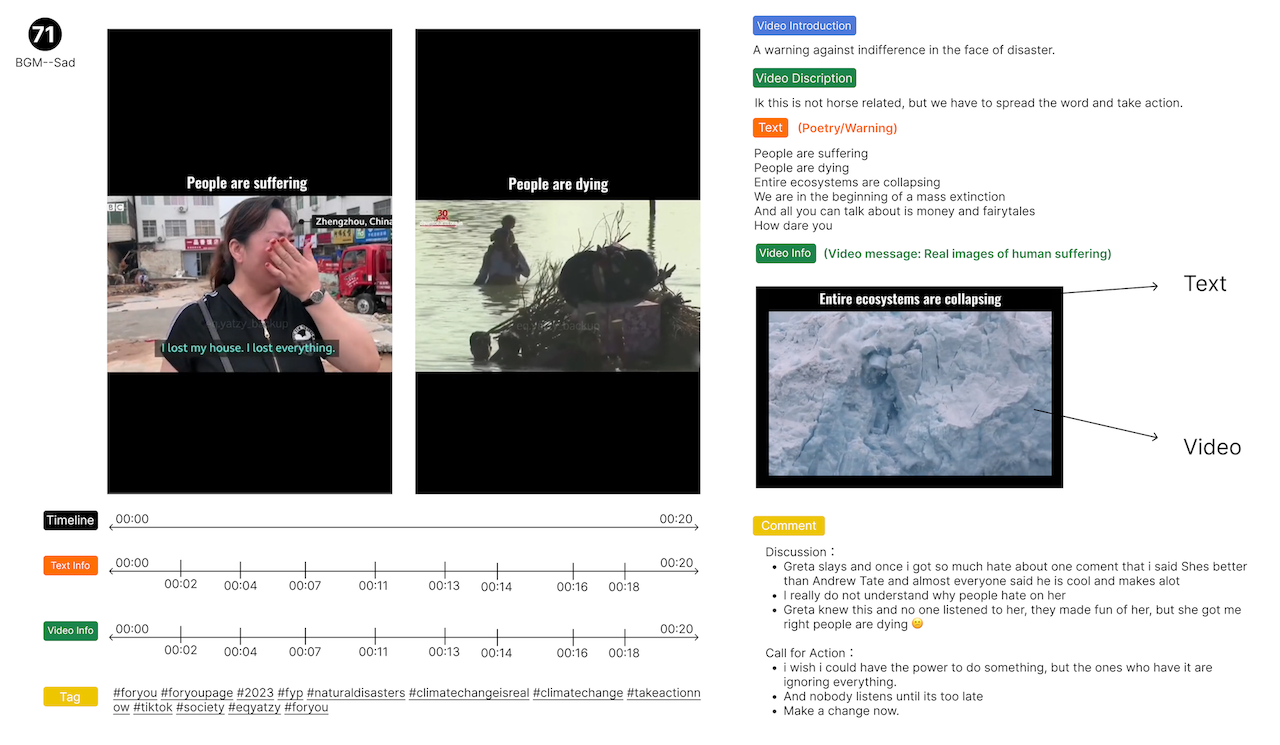}
   \caption{An example of content analysis of climate change videos}
    \label{fig1} 
\end{figure}

We implemented a systematic coding framework to document: (1) content elements (narrative summaries, representative screenshots), (2) technical features (background music, duration), (3) textual components (descriptions, tags), and (4) timestamped visual elements (on-screen text overlays, image insertions, third-party footage integration), and (5) audience response patterns (summary of comment characteristics, focusing on creator-viewer interaction, primarily using the \textcolor{blue}{top 6} most liked comments). An example of this analytical framework is presented in Figure \ref{fig1}.

\setlength{\textfloatsep}{4pt}

Based on this systematic coding framework, each of the four researchers was assigned 25 videos for analysis. To ensure consistency and reliability of the analytical results, we held regular review meetings to cross-check and complement one another's analyses. For example, when discrepancies in interpretation or inadequate reflections of the video's main theme were identified during discussions, we would collaboratively refine the analytical framework until consensus was reached. This mechanism ensured an accurate understanding of the video content and types. Finally, we visualized the expression strategies for each video type.

\subsubsection{Analysis of Viewers Responses}\

\setlength{\textfloatsep}{5pt}
\begin{figure}[t]
    \centering
\includegraphics[width=1\linewidth]{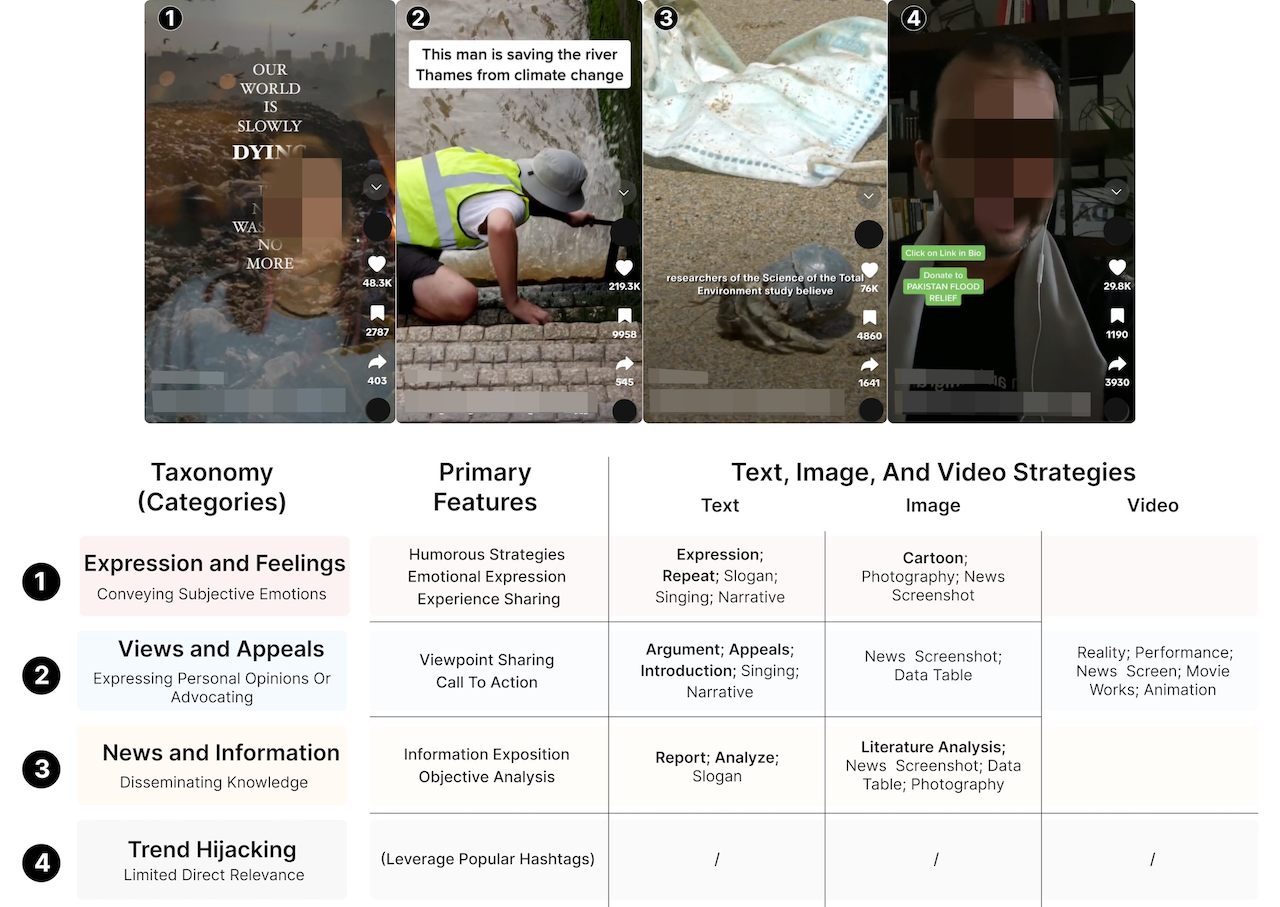}
    \caption{Key features and multimedia strategies for each video category: 1) \textit{Expression and Feelings}, 2) \textit{Views and Appeals}, 3) \textit{News and Information}, and 4) \textit{Trend Hijacking}}
    \label{fig2}
\end{figure}

While analyzing the video content, we conducted a qualitative analysis of audience comments. Since our analysis is based on climate communication theory, which emphasizes the importance of communication in climate discourse, we specifically focused on creator-viewer interaction. Additionally, to ensure that the comments section reflects the viewpoints of the majority, we selected the top 6 comments~\cite{creswell2017research}, as these typically represent the dominant perspectives within the audience. In cases where a video's comment section lacked creator-viewer interaction, we proceeded by selecting the top 6 most liked comments to maintain a consistent sample size across all videos.

After conducting qualitative analysis of the comments in the \textcolor{blue}{top 6} and those with interaction, we extracted 1–2 keywords to summarize the comment trends for each video (this definition, as mentioned in the previous chapter, was discussed and agreed upon with the research team on a regular basis). Subsequently, we presented interaction case studies for each video type, and calculated and compared the frequency of these keywords across different video types to identify audience response trends. This analysis helps reveal how different climate communication strategies elicit audience reactions.

\section{Results}\label{sec:Results}
\subsection{Categorizing Climate Change Short Videos on TikTok}
\label{sec:4.1}

\setlength{\textfloatsep}{5pt}
\begin{figure}[t]
    \centering
    \includegraphics[width=0.7\textwidth]{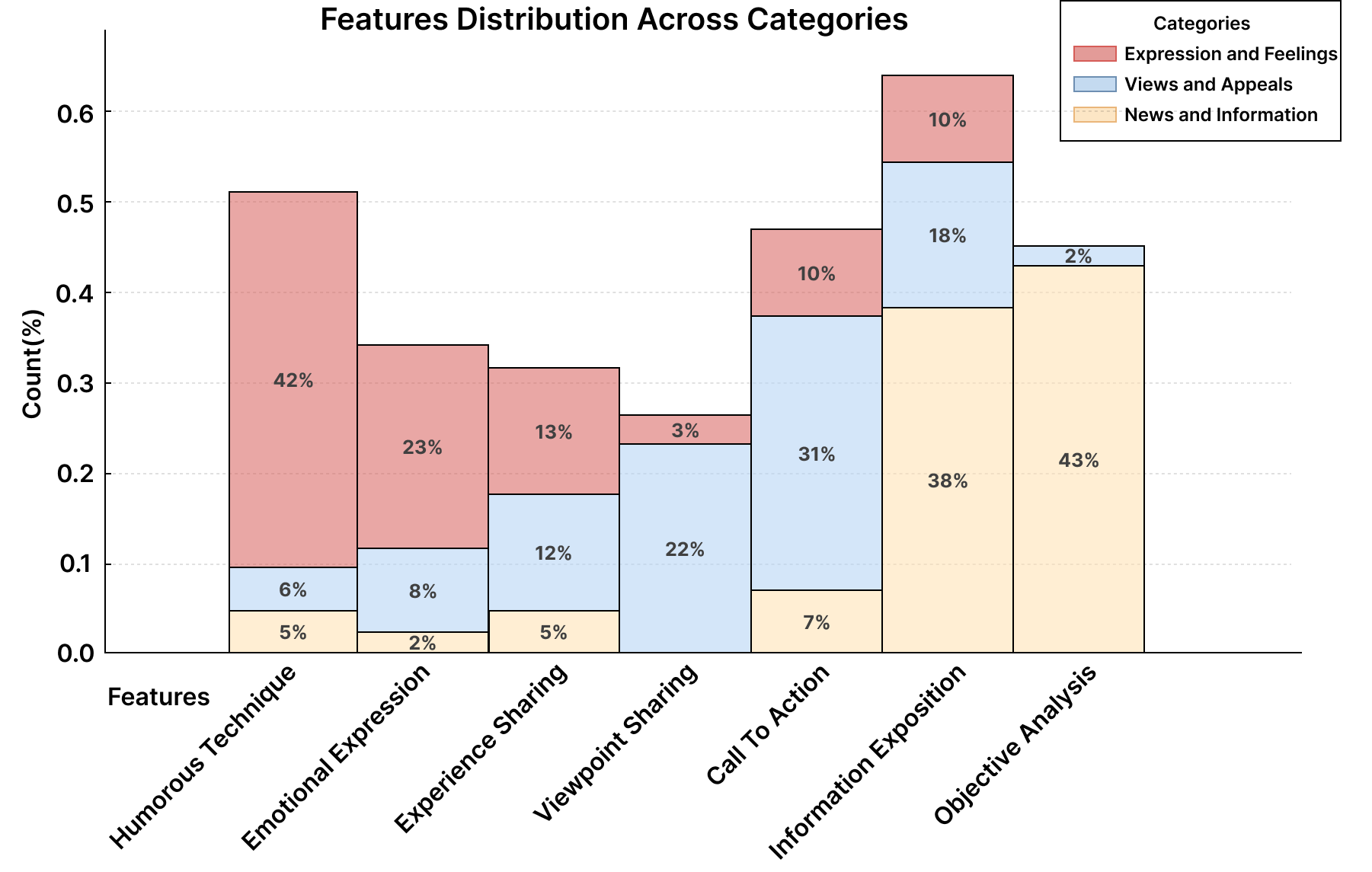}
    \caption{Stylistic and technical features distribution across video categories: The primary features of \textit{Expression and Feelings} videos are a. Humorous Strategies (using protest imagery cautiously), b. Emotional Expression (showcasing genuine emotions and images of people), and c. Experience Sharing (telling new stories about climate change). \textit{Views and Appeals} videos mainly emphasize d. Viewpoint Sharing (communicating with the viewers) and e. Call to Action (highlighting the impacts of climate change). \textit{News and Information} videos are dominated by f. Information Exposition (presenting local impacts of climate change) and g. Objective Analysis (illustrating the scale and specific causes of climate change).}
    \label{fig3}
\end{figure}

This section presents our qualitative analysis of a random sample of 200 TikTok videos (Figure \ref{fig3}) and provides a detailed description of each category of video we identified (Figure \ref{fig2}). \textit{Expression and Feelings} (21.5\%) videos focus on conveying subjective emotions, including straightforward expressions such as sadness, anger, and excitement, as well as more nuanced emotions like irony, resignation, and humor. For example, they may use clips of real-life scenes to highlight the planet's environmental decline. \textit{Views and Appeals} (30.5\%) videos concentrate on expressing personal opinions or social advocacy, often showcasing individual actions to underscore the urgent need for transformation. \textit{News and Information} (28\%) videos are dedicated to disseminating knowledge about climate change, using news footage or personal narration supported by authoritative charts to highlight the objectivity of the information. For instance, they may report on the staggering scale of coastal debris to stress the issue's severity. \textit{Trend Hijacking} (20\%) videos have limited direct relevance to climate change but use related hashtags to increase visibility. Their content spans topics such as politics, war, commercial promotions, and everyday life. These four categories each reflect the diverse strategies users employ in their expression, focusing respectively on emotional conveyance, opinion advocacy, information dissemination, and trend exploitation.

\subsubsection{Expression and Feelings}\

\setlength{\textfloatsep}{5pt}
\begin{figure}[t]
    \centering
    \includegraphics[width=1\textwidth]{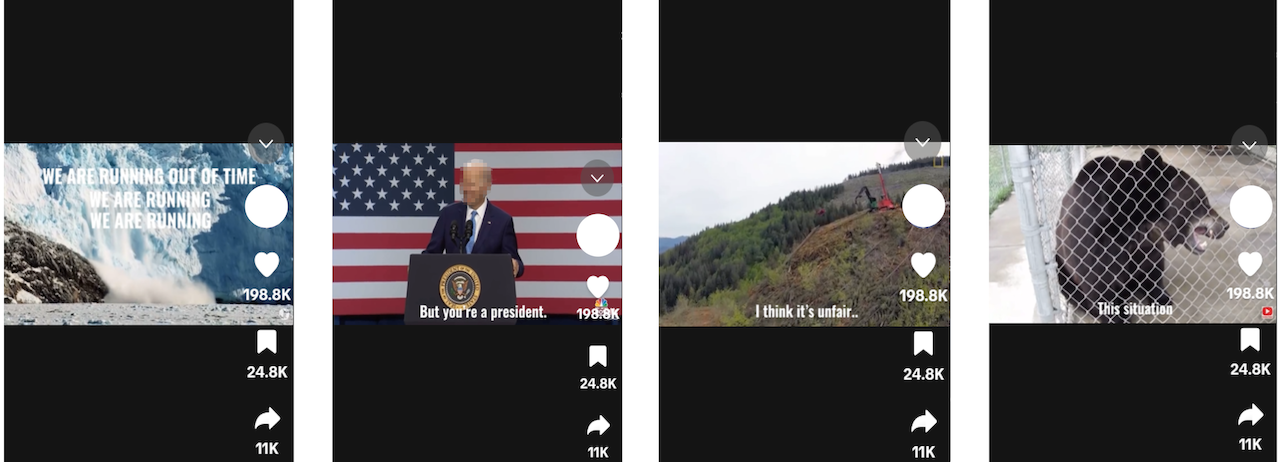}
   \caption{Humorous strategies: The creator has changed the content of the lyrics through subtitles, matching the \textit{president} with the \textit{killer} and \textit{random teenager} for the \textit{best friend}, highlighting the creator's satire on specific political environmental measures. The other negative images displayed in the video also correspond to the lyrics.}
    \label{fig4}
\end{figure}

The most prominent feature of \textit{Expression and Feelings} videos is ``Humorous Strategies'' (42\%). These videos express emotional reactions to climate issues in lighthearted or humorous ways, aiming to attract audiences to climate topics in a non-serious manner. Satire and irony are often used to mock the actions of governments or public figures, rather than directly using protest visuals. For example, \begin{quote}
    \textit{One video uses rewritten song lyrics to critique a specific policy, pairing the altered lyrics with imagery of affected wildlife to underscore the critique's emotional weight} (see Figure \ref{fig4}).
\end{quote} Other common techniques include puns, dark humor, stand-up comedy clips, meme-style edits, drawings, and skits. These entertaining forms strike a balance between conveying information and providing amusement. In addition, creators also use sound effects, visual elements, and humorous animations to enhance appeal, attempting to embed serious information through clever text and dialogue, thereby prompting audiences to reflect deeply on the severity of climate change. These formats aim to attract viewers to climate topics in an accessible manner.

\setlength{\textfloatsep}{5pt}
\begin{figure}[t]
    \centering
    \includegraphics[width=1\textwidth]{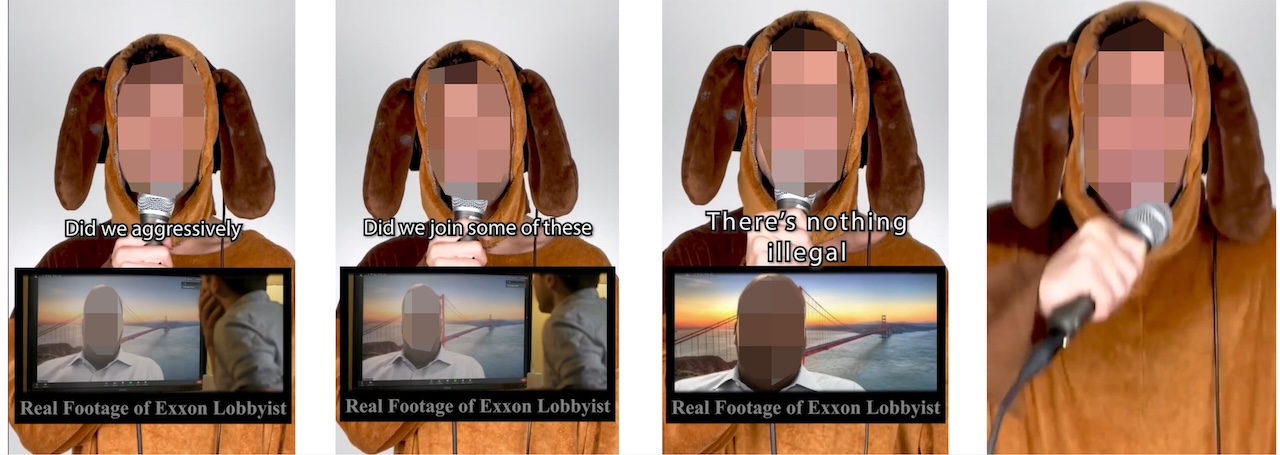}
   \caption{Emotional expression: The image shows the creator using music and rap to shout out opinions angrily.}
    \label{fig7}
\end{figure}

The second most prominent feature of \textit{Expression and Feelings} videos is ``Emotional Expression'' (23\%). \textit{Expression and Feelings} videos excel in conveying strong emotions such as anger, frustration, or sadness through intense personal reactions (such as loud shouting or grimacing). These videos are also accompanied by visually impactful imagery (e.g., rivers filled with plastic bottles, wildfires, emaciated polar bears) and moving narratives (e.g., recounting the sorrow of witnessing deforestation and urging protection for future generations). \begin{quote}
    \textit{As shown in Figure \ref{fig7}, a creator combines music and rap with angry outbursts to express dissatisfaction with policy decisions. }
\end{quote}The strategic combination of intense personal outbursts and high-impact visual imagery shifts the focus from abstract global warming to visceral, localized distress, effectively transforming climate change into a shared emotional crisis rather than a distant scientific fact.

\setlength{\textfloatsep}{5pt}
\begin{figure}[t]
    \centering
    \includegraphics[width=1\textwidth]{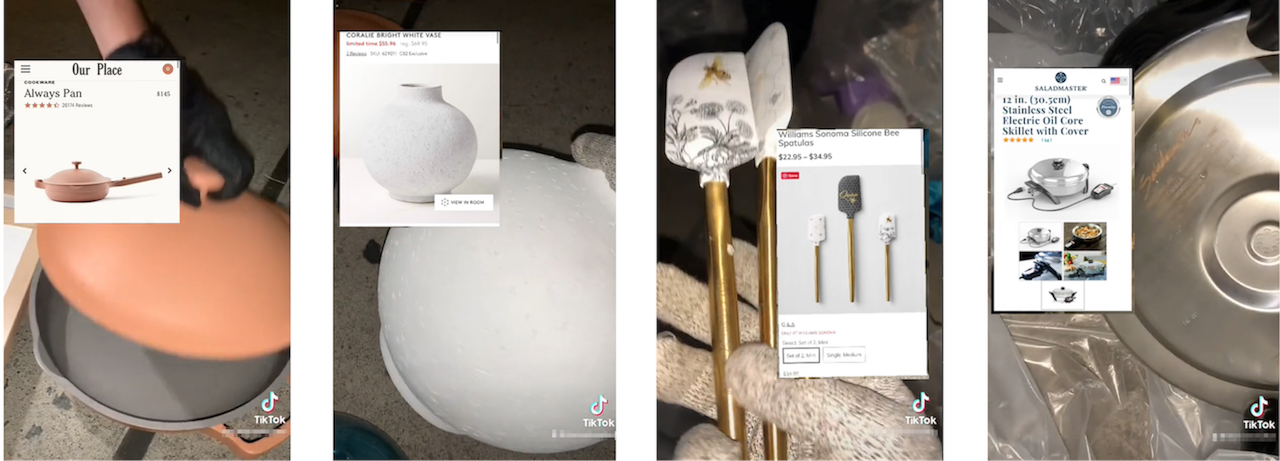}
   \caption{Experience sharing: The image shows the creator shares her personal experience of picking up garbage.}
    \label{fig8}
\end{figure}

The third most prominent feature in \textit{Expression and Feelings} videos is ``Experience Sharing'' (13\%). These videos influence people's perceptions through personal stories or shared experiences. \begin{quote}
    \textit{Figure \ref{fig8} illustrates this with an example of a video documenting real community cleanup efforts. These videos not only convey personal experiences but also encourage broader discussions about climate change.}
\end{quote} In contrast, only 5\% of \textit{News and Information} videos exhibit this feature. The difference is that news content focuses more on factual reporting and information delivery rather than emotional expression or personal experience. This emphasis on lived experience suggests that local participation provides a more relatable entry point for audiences.

\subsubsection{Views and Appeals}\

\begin{figure}[t]
    \centering
    \includegraphics[width=1\textwidth]{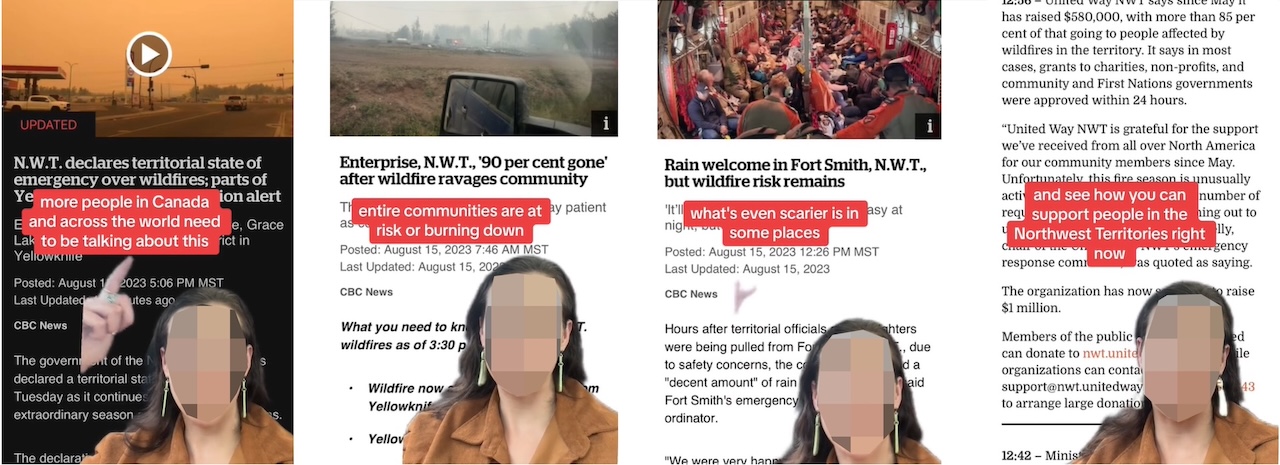}
   \caption{Call to action: The image shows the creator calling for attention to and protection of wildlife.}
    \label{fig9}
\end{figure}

\textit{Views and Appeals} videos' most prominent feature is ``Call to Action'' (31\%), which appears three times more frequently than in the other two categories. This phenomenon can be attributed to the nature of such content, which frames climate change as an urgent collective challenge, emphasizing its impact on daily life and encouraging specific actions in response. These videos often include explicit calls to action, such as sharing personal efforts (e.g., deleting spam emails), advocating policy changes (e.g., reducing electricity bills), or promoting environmentally friendly practices (e.g., adopting low-carbon travel or reducing meat consumption). These calls are closely tied to everyday life, making them easy to understand and highlighting the personal issues behind public engagement with climate change. \begin{quote}
    \textit{As shown in Figure \ref{fig9}, one creator utilizes a recent environmental crisis to urge audience involvement, accompanied by on-screen text with a clear instruction for communal support.}
\end{quote} By translating grand environmental narratives into actionable micro-steps, creators reduce the psychological distance of climate change, making a global crisis feel manageable at an individual level.

\setlength{\textfloatsep}{5pt}

\begin{figure}[t]
    \centering
    \includegraphics[width=1\textwidth]{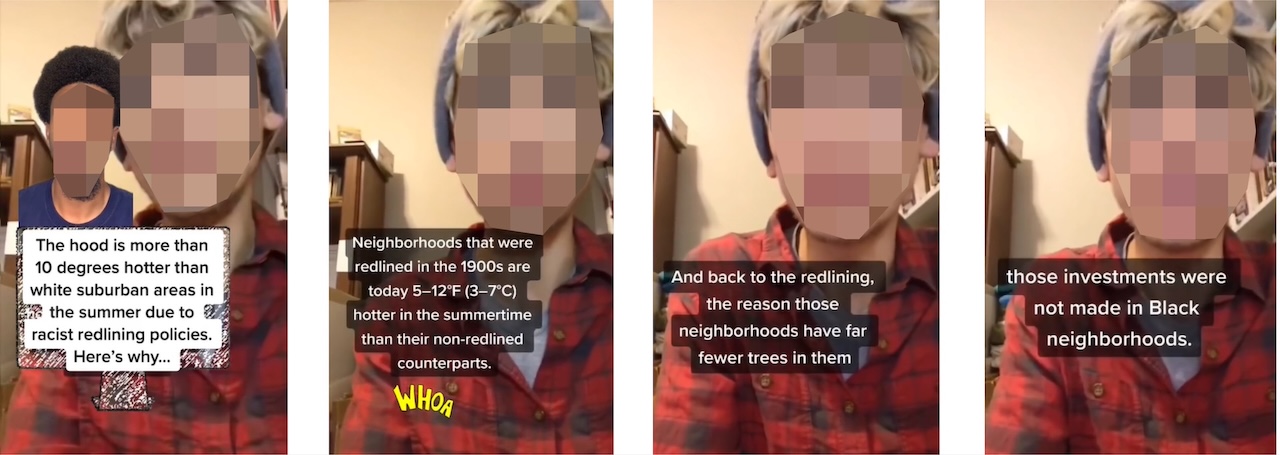}
   \caption{Viewpoint sharing: The image shows the creator thinks climate justice is about racial justice.}
    \label{fig10}
\end{figure}

\begin{figure}[t]
    \centering
    \includegraphics[width=1\textwidth]{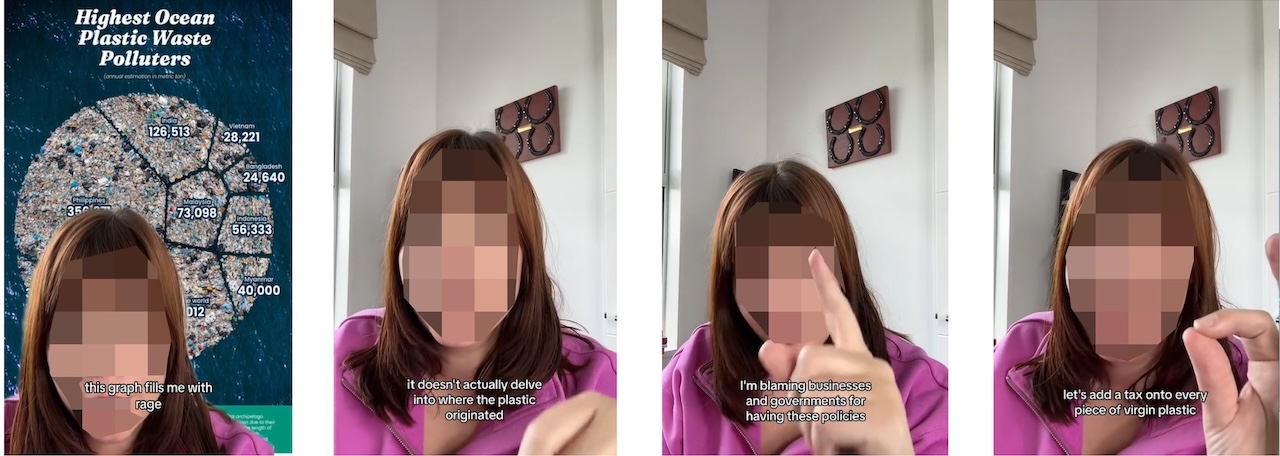}
   \caption{Viewpoint sharing: The first half of the video took an objective analysis of plastic pollution and ultimately output a subjective blame of personal frustration and a call for the content of fiscal policy implementation.}
    \label{fig11}
\end{figure}

The second most prominent feature in \textit{Views and Appeals} videos is ``Viewpoint Sharing'', accounting for 22\% of the total. These videos focus on sharing personal viewpoints with the audience and emphasizing critical thinking. Compared to factual content, these videos place more emphasis on subjective expression. For example,  \begin{quote}
    \textit{As shown in Figure \ref{fig10}, a creator links climate issues to racial justice,} offering a perspective that challenges mainstream narratives and invites the audience to engage in discussions about differing viewpoints. 
\end{quote}This intersectional approach suggests that some TikTok creators view climate change as a socio-political catalyst for social justice, rather than an isolated scientific issue. Since personal opinions lack the rigor and validation of authoritative sources, even when data visualizations are included, they are often used to support subjective views rather than reflect objective facts. \begin{quote}
    \textit{As shown in Figure \ref{fig11}, the creator criticizes institutional stakeholders regarding pollution and calls for specific economic interventions.}
\end{quote}Although official data lists the issue of ocean plastic pollution, the opinions and calls made in the video are highly subjective, demonstrating a strategic use of data where quantitative evidence is repurposed to amplify moral indignation rather than merely to inform.

\subsubsection{News and Information}\

\begin{figure}[t]
    \centering
    \includegraphics[width=1\textwidth]{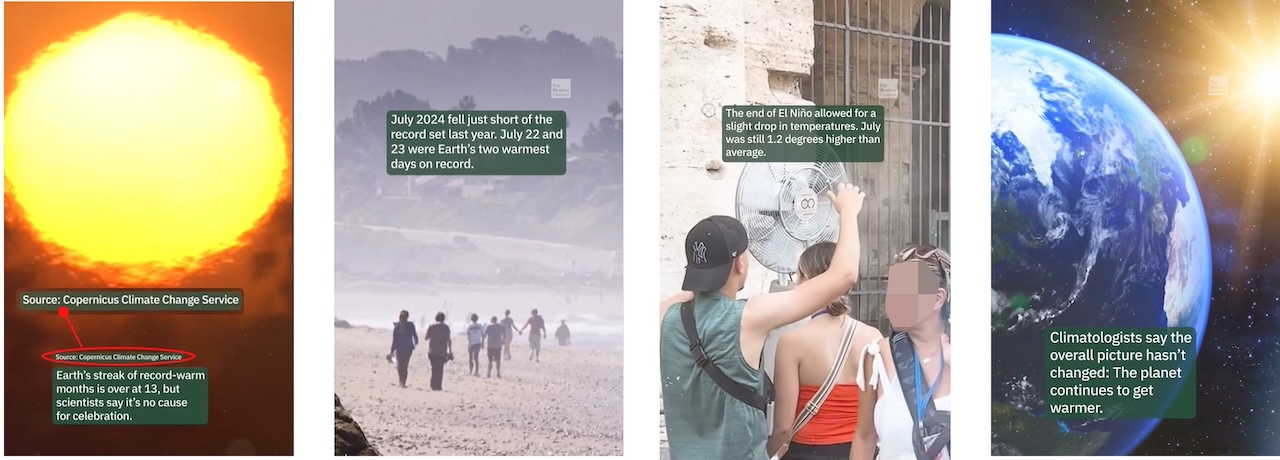}
   \caption{Objective analysis: The image shows the current state of climate warming worldwide.}
    \label{fig14}
\end{figure}

\textit{News and Information} videos' most prominent feature is ``Objective Analysis'' (43\%). The broad appeal of Objective Analysis videos lies in their fact-based approach. These videos aim to provide objective, data-driven insights about climate change, particularly expert opinions. The primary goal of Objective Analysis videos is to explain the complexities of climate change, including its causes, impacts, and other related aspects. For example,\begin{quote}
    \textit{A video presents data on regional heatwaves, reporting on the severity of recent weather conditions and emphasizing the recorded temperature extreme (see Figure \ref{fig14}).}
\end{quote} These videos aim to provide an impartial review of scientific evidence, policy impacts, and environmental statistics, demonstrating the concrete facts of climate change and encouraging audiences to think deeply about its implications. This emphasis on impartiality provides a sense of reliability within the fast-paced TikTok feed, offering a factual balance to the emotional content commonly found on the platform.

\begin{figure}[t]
    \centering
    \includegraphics[width=1\textwidth]{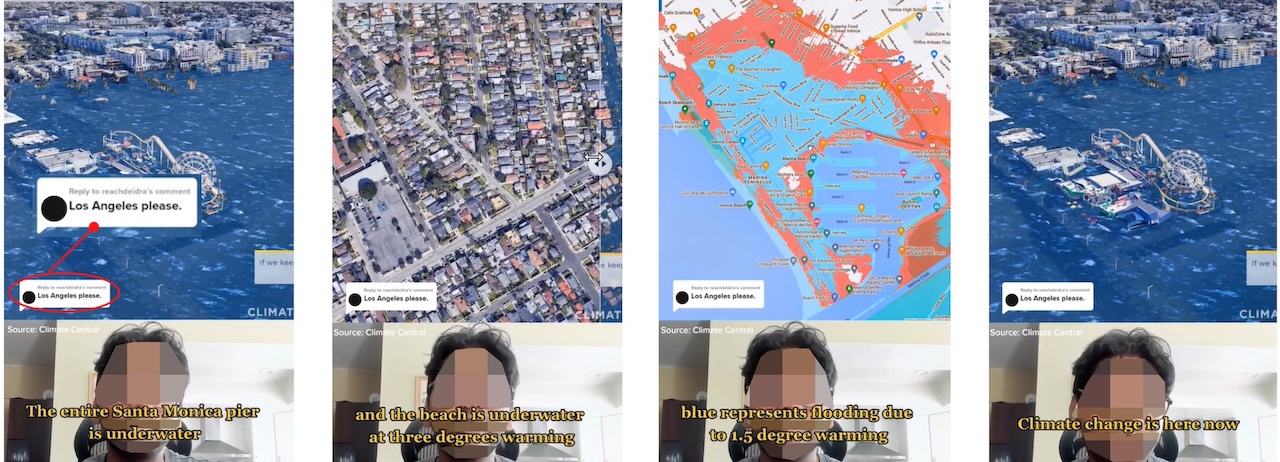}
   \caption{Information exposition: The analysis of the current sea level rise.}
    \label{fig12}
\end{figure}

\begin{figure}[t]
    \centering
    \includegraphics[width=1\textwidth]{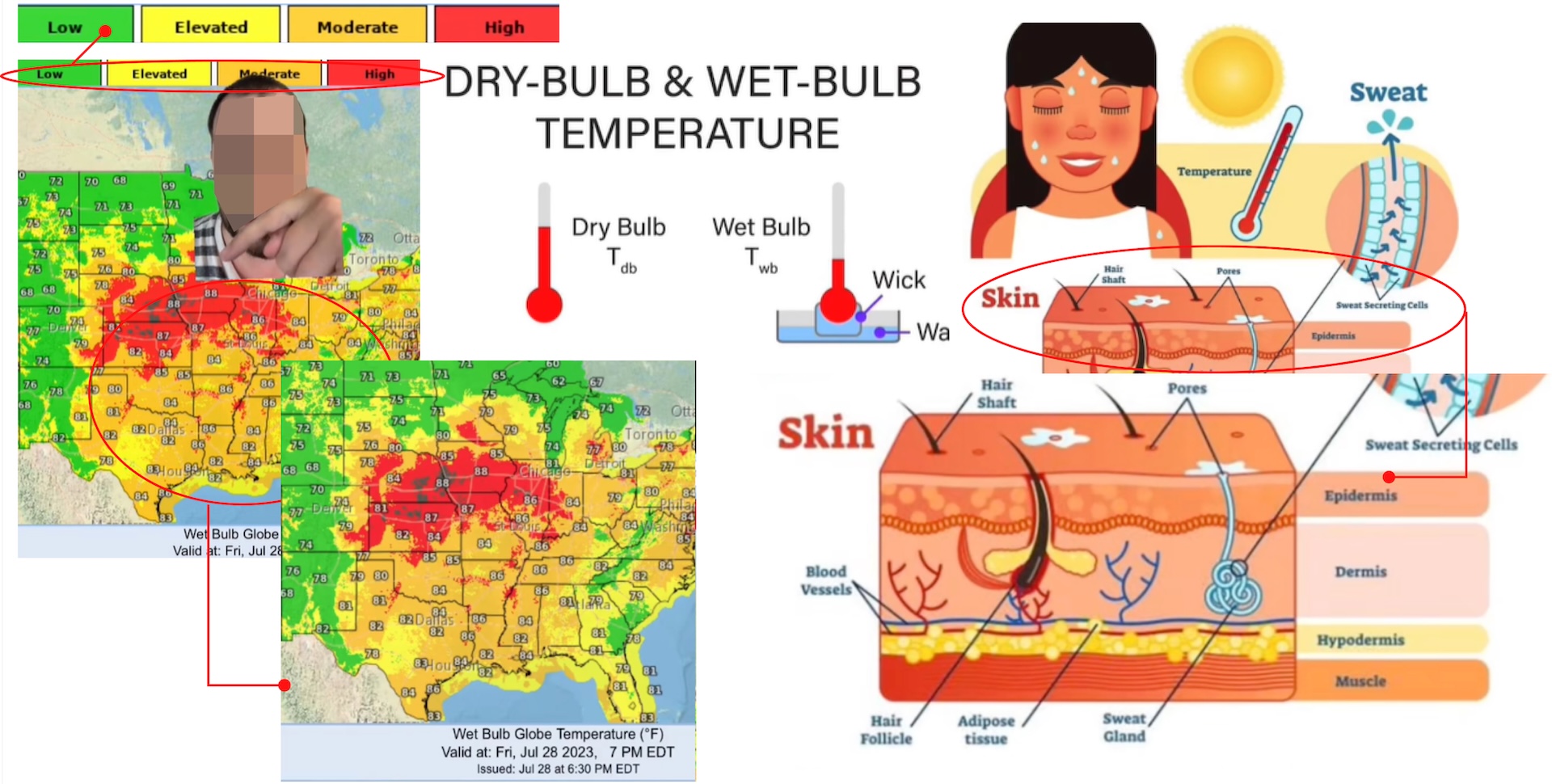}
   \caption{Information exposition: A scientific explanation of temperature trends in a North American country using heat maps, scientific charts, and biological illustrations.}
    \label{fig13}
\end{figure}

The second most prominent feature of \textit{News and Information} videos is ``Information Exposition'' (38\%). Although all three video categories contain this feature, it is most prevalent in \textit{News and Information} videos. This prominence can be attributed to the nature of \textit{News and Information} content, which inherently focuses on providing local facts and educational information about climate change. These videos often present on-site footage, data (e.g., pie charts, heat maps), scientific findings, and expert opinions in the form of objective reports. For example, \begin{quote}
    \textit{A creator uses numerical data and scientific evidence to illustrate the current state of sea levels (see Figure \ref{fig12}).}
\end{quote} These videos deepen audiences' understanding of climate change issues by providing science-based content. They aim to help the public gain meaningful knowledge through clear explanations and reliable information, encouraging informed participation in climate action. \begin{quote}
    \textit{As shown in Figure \ref{fig13}, the video creator uses heat maps, scientific charts, and biological illustrations to explain regional temperature changes and the causes behind the temperature drops.}
\end{quote} This pedagogical approach suggests that TikTok is evolving beyond mere entertainment into an important platform for informal science learning, where the visual simplification of regional temperature changes helps bridge the gap between expert discourse and public understanding.

\subsubsection{Trend Hijacking}\

\begin{figure}[t]
    \centering
    \includegraphics[width=1\textwidth]{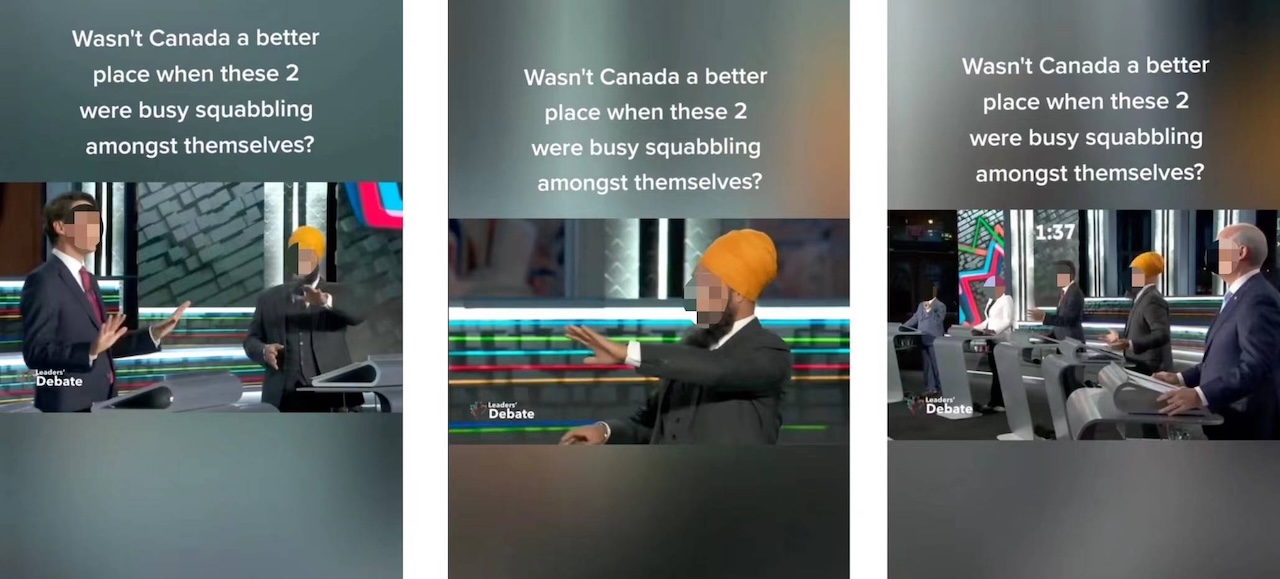}
   \caption{Screenshots from a video of two politicians debating current affairs on a TV show}
    \label{fig26} 
\end{figure}

\begin{figure}[t]
    \centering
    \includegraphics[width=1\textwidth]{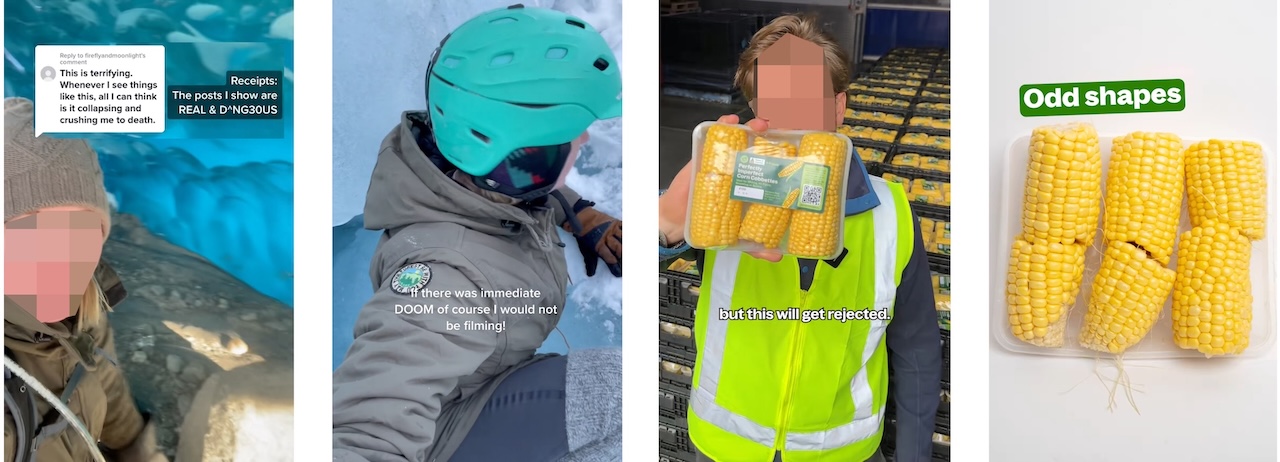}
   \caption{Left (a-b): Screenshots from a glacier exploration video; Right (c-d): Screenshots from a corn purchase video}
    \label{fig28} 
\end{figure}

The feature of \textit{Trend Hijacking} videos lies in using hashtags related to social hot topics to attract traffic, covering a wide range of themes, including international and social issues as well as daily life sharing. \textcolor{blue}{Content creators} leverage popular hashtags not only to increase exposure but also to capitalize on trends to capture the attention of viewers. For example, \begin{quote}
    \textit{One video incorporates a segment from a televised political debate focusing on a contentious policy issue (see Figure \ref{fig26}). }
\end{quote}The video uses 92 hashtags, including \#climatechange, aiming to attract public attention to this popular topic. The \textit{Trend Hijacking} effect is quite significant, with the video garnering 867,567 views and 1,298 comments.

Additionally, some videos address topics such as consumer perceptions of food aesthetics, encouraging viewers to look beyond superficial qualities and appreciate other attributes of produce. Other videos respond to fans' questions about glacier exploration, providing related suggestions and precautions (see Figure \ref{fig28}). Through their broad selection of topics and interaction with audiences, these videos not only entertain but also serve broader purposes, such as raising awareness, promoting ideas, or issuing public warnings, all while leveraging trending hashtags to amplify their reach.

\subsection{Analysis of Visual, Textual, and Auditory Strategies by Video Categories}
\label{sec:4.2}

To understand the differences in expression strategies between these video categories, we conducted a detailed content analysis of 25 videos from each category, examining characteristics such as expressive techniques, types of information, and emotional tendencies in climate change-related videos on TikTok.

\setlength{\textfloatsep}{5pt}
\begin{figure}[t]
    \centering
    \includegraphics[width=1\linewidth]{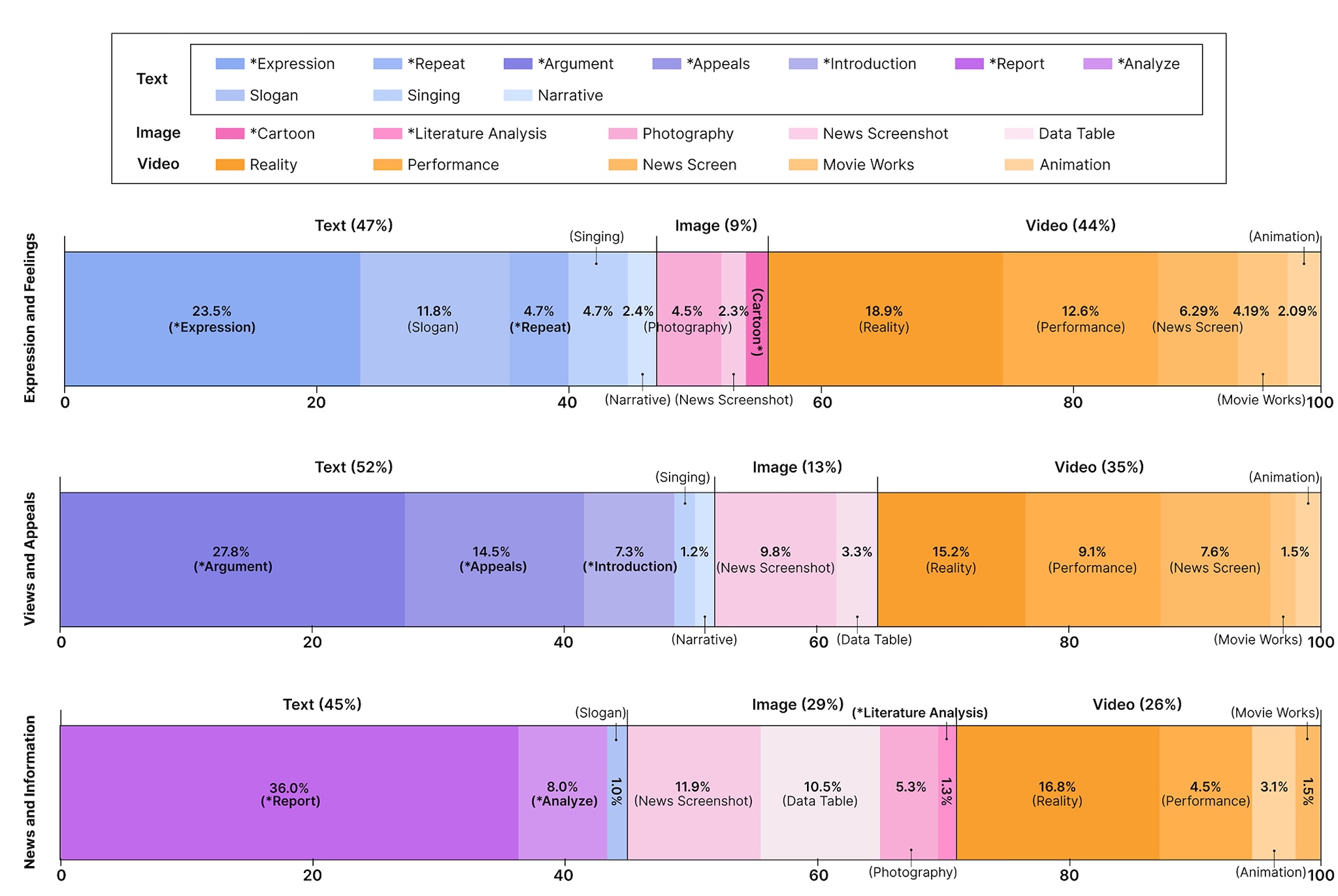}
    \caption{Distribution of text, image, and video strategies across categories (``*'' indicates that these strategies only appear in specific video categories)}
    \label{fig15}
\end{figure}

The distribution of media elements—such as text, images, and videos—reveals how different video categories shape their distinct styles and communicate information effectively (Figure \ref{fig15}). Combining element analysis with case studies reveals how various strategies shape emotional, aesthetic, and informational impacts across different video categories. \textit{Expression and Feelings} videos primarily use everyday life scenes to convey personal climate feelings. A key distinction is their use of expressive, repetitive text to convey information. Given the humorous tone of these videos, cartoon-style imagery is also frequently incorporated. This combination of content and presentation effectively evokes the fluctuations of emotional expression. \textit{Views and Appeals} videos tend to contain more text-based content, using arguments and appeals to communicate messages. Emphasizing text helps convey ideas more clearly and encourages engagement through calls to action. \textit{News and Information} videos use balanced media elements to reflect data richness and maintain objectivity. Text in these videos emphasizes reports, analyses, and other factual content, often supplemented by news screenshots, datasets, or research literature to underscore objectivity. This balanced media use reinforces the reliability and authority of the information.

\subsubsection{Emotional Expression Strategies in Expression and Feelings Videos}\

\setlength{\textfloatsep}{5pt}
\begin{figure}[t]
    \centering
    \includegraphics[width=1\linewidth]{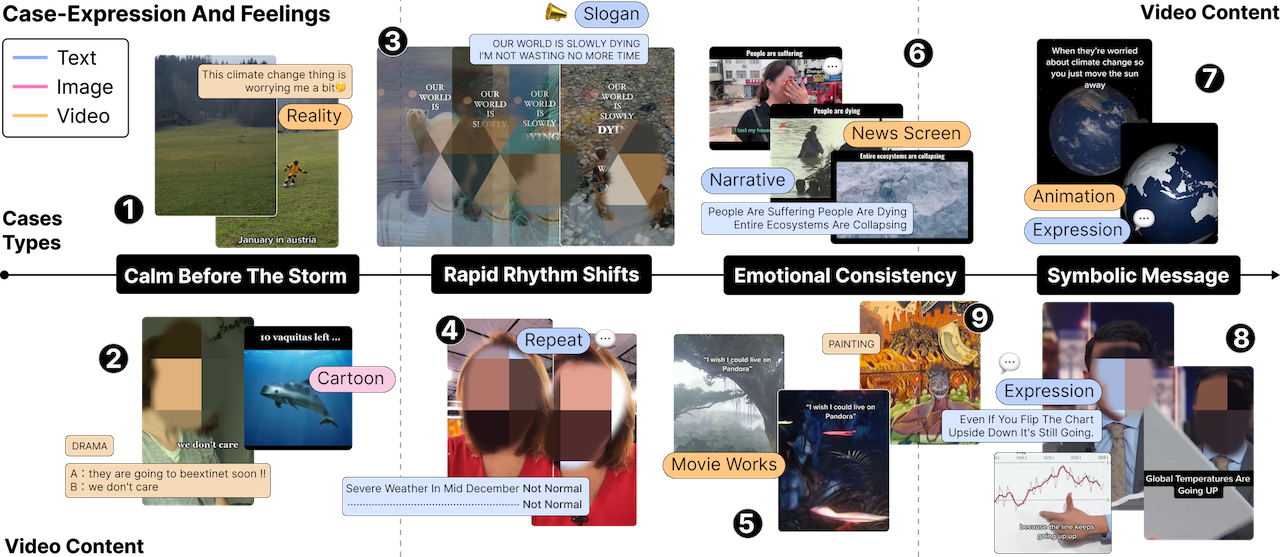}
    \caption{The trend of emotional expression in \textit{Expression and Feelings} and cases of video content}
    \label{fig17}
\end{figure}

In \textit{Expression and Feelings} videos, emotional expression is often dominated by negative emotions, such as sadness, anger, panic, and helplessness (Figure \ref{fig17}). The expression of emotions in these videos goes beyond simple descriptions, employing a combination of various elements to enhance immersion. For example,
\begin{quote}
    \textit{One video opens with a person skiing in a beautiful natural setting (Figure \ref{fig17}(1)), accompanied by soft background music and serene visuals, creating a sense of tranquility.}
\end{quote} This mood is subtly undercut by an on-screen caption expressing concern about climate change. As viewers immerse themselves in the peaceful atmosphere, the story suddenly turns unexpectedly—just as the caption hints—and the skier falls due to insufficient snow. This sudden reversal breaks the peaceful mood, using visual and emotional conflict to emphasize the problems posed by climate change. Similar techniques are used in a dialogue sequence.
\begin{quote}
    \textit{The exchange presents a dismissive response to a statement regarding species endangerment, followed by an abrupt cut to a cartoon of the affected animal alongside stark numerical information regarding its remaining population. This transition is accentuated by unsettling background music (see Figure \ref{fig17}(2)).}
\end{quote} This absurd contrast reinforces the satirical tone, allowing the topic to be conveyed in a playful atmosphere. Creators help audiences process overwhelming climate anxiety without feeling immediately defeated.

Rapid changes in rhythm are also common in this video type, with videos often building tension through silence or gradually intensifying background music. For example, \begin{quote}
    \textit{One video begins with a silent and disappointed headshake (Figure \ref{fig17}(3)), which is then punctuated by a rapid montage of images depicting environmental degradation.}\end{quote} The on-screen text is dynamically unveiled, beginning with a fragmentary phrase that culminates in a complete and grave declaration about the slogan of the world, thereby heightening both the visual rhythm and rhetorical impact. Some videos use repetition to amplify emotional expression, \begin{quote}
    \textit{Such as repeating phrases that highlight perceived abnormality (Figure \ref{fig17}(4)), often with rising vocal intensity to convey the anxiety surrounding extreme weather events.}\end{quote} The combination of repetitive language and escalating tone creates a sense of urgency. Some videos abruptly end to maintain emotional tension, while others slow down with soothing music (Figure \ref{fig17}(5)). This careful control of pacing creates a balance between emotional intensity and relaxation. Such rhythmic variations prevent message fatigue, keeping audiences engaged with repetitive environmental themes by constantly shifting the emotional pressure.

However, even within this calmness, subtle fluctuations and underlying tension remain noticeable. In contrast, some narrative-style videos take a gentler approach, juxtaposing beautiful natural landscapes with distressing scenes (Figure \ref{fig17}(6)). Although negative emotions are interspersed throughout, the overall tone remains more moderate, focusing on storytelling. For example, \begin{quote}
    \textit{A video uses a sequence of stark, declarative statements as on-screen text to outline a progression of ecological and social crises, using narrative elements to convey the gravity of the situation while encouraging calm reflection (Figure \ref{fig17}(5)).}\end{quote} In some instances, emotional expression is conveyed indirectly through metaphors. \begin{quote}
    \textit{An animation portrays a cold Earth after losing the Sun, accompanied by the ironic caption that suggests removing the sun as a tongue-in-cheek solution to climate concerns (Figure \ref{fig17}(7)).}\end{quote} This metaphor conveys serious messages in an engaging context, lowering the barrier to audience understanding. Similarly, humor is used creatively, \begin{quote}
    \textit{Such as in a stand-up comedy routine that exaggerates data to convey the severity of climate change by comically asserting that a trend would continue unabated even if its supporting chart were inverted (Figure \ref{fig17}(8)).}\end{quote}\begin{quote} Another video uses art to depict the state of the Earth,
    \textit{With an artist immersively illustrating their imagined vision of the planet's condition (Figure \ref{fig17}(9)).}\end{quote} These creative forms show that TikTok provides a space for reinterpreting complex scientific crises through personal and artistic lenses, making it easier for all viewers to engage deeply, whether or not they were previously concerned about climate change.

In summary, \textit{Expression and Feelings} videos utilize the interplay of various elements to convey emotions in a layered and nuanced way. They emphasize contrast and rhythm, combining visuals, sound, and narrative to deepen emotional engagement. The creators of these videos attempt to evoke emotional responses in a short time, using artistic expression to resonate with audiences and provide an effective means of communicating serious social issues.

\subsubsection{Persuasion Through Narrative Strategies in Views and Appeals Videos}\

In \textit{Views and Appeals} videos, narratives serve not only as a means of conveying viewpoints but also as a way to create a dual connection between storytelling and emotions through the use of various elements. These videos often focus on opinions about extreme weather events or commentary on news stories, following two primary approaches: one emphasizes data and phenomena analysis, highlighting the need for solutions, while the other illustrates potential catastrophic consequences, urging viewers to take action to prevent future losses.

\begin{figure}[t]
    \centering
    \includegraphics[width=1\linewidth]{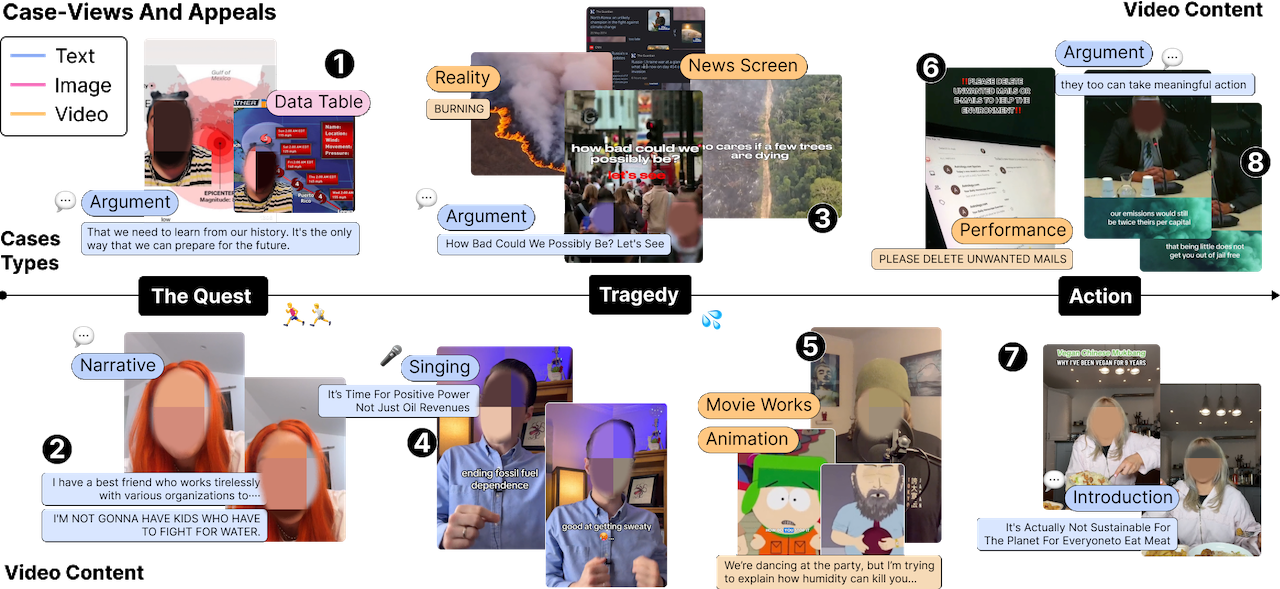}
    \caption{The trend of storytelling in \textit{Views and Appeals} and cases of video content}
    \label{fig19}
\end{figure}

The first approach often integrates news clips, screenshots, and collections of key events to build coherent information. This synthesis of disparate media elements allows creators to construct a self-contained evidence base, effectively mimicking journalistic practices to lend an air of investigative credibility to their personal viewpoints. For example,\begin{quote}
    \textit{One video uses screenshots of typhoons and earthquakes occurring in the same month and date (Figure \ref{fig19}(1)).}\end{quote} The accompanying on-screen text advances a viewpoint on the cyclical nature of historical events, suggesting that patterns from the past are poised to recur. This expression leverages historical coincidences, using news visuals and text to emphasize the necessity of learning from the past as a crucial means of preparing for what lies ahead. Creators attempt to enhance the credibility of their viewpoints by using repeated news images while also conveying a sense of urgency to the audience. Other videos feature personal speeches as their narrative core, \begin{quote}
    \textit{Such as highlighting efforts to protect Indigenous communities. The oral delivery is reinforced by captions that convey a resolve to shield future generations from resource-based conflicts, particularly over essentials such as water (Figure \ref{fig19}(2)).}\end{quote} The progression of subtitles and the speaker's tone convey the urgency of the climate crisis, culminating in a spoken appeal that urges viewers to confront their fear and translate it into collective action. Creators attempt to engage viewers more directly with the interplay of emotions and information by combining narration with sound, text, and subtitles.

\setlength{\textfloatsep}{5pt}

The second narrative approach focuses on the significant losses and disasters that may occur if no action is taken. These videos not only emphasize the severity of the problem from a natural perspective but also explore the risks of climate change from a human standpoint. For example, \begin{quote}
    \textit{One video switches between scenes of wildfires, landslides, and floods. The on-screen text poses a rhetorical question about the extent of human impact, followed by a prompt inviting the viewer to witness the presented evidence (Figure \ref{fig19}(3)).}\end{quote} The alternation between catastrophic visuals and concise slogans creates a powerful visual impact, allowing viewers to grasp the destructive force of climate change intuitively. Other videos explore the impact of humidity on human health. \begin{quote}
    \textit{As shown in Figure \ref{fig19}(4), presenting scientific information regarding the hazardous conditions that can arise when specific temperature thresholds are accompanied by extreme levels of humidity.}\end{quote} Creators warn about the health risks associated with rising temperatures and humidity, advocating for a fundamental reduction in societal reliance on the combustion of fossil fuels. \begin{quote}
    \textit{Another video humorously references a well-known animated series to explain the relationship between humidity and temperature (Figure \ref{fig19}(5)).}\end{quote} The narration uses a lighthearted scenario of a social gathering to introduce a serious explanation of how elevated humidity levels can pose lethal risks to human physiology, particularly by impeding the body's natural cooling mechanism. The video transitions from a playful party scene to captions analyzing humidity and health risks, with sound effects matching the tonal shifts. The creator combines visuals, text, and sound to interact seamlessly, not only conveying information but also encouraging viewers to actively share knowledge and advocate for concrete measures to be taken. These creators attempt to convey serious messages through light-hearted language, making it easier for audiences to understand and accept the gravity of climate change.

Many videos also feature personal actions to inspire viewers to practice environmental responsibility. For example,\begin{quote}\textit{One video shows someone deleting spam emails with the caption (Figure \ref{fig19}(6)) that explicitly calls viewers to delete unwanted emails as a way to help the environment by reducing e-waste.}\end{quote}
Another video shows a person eating a vegetarian meal.
\begin{quote}\textit{As shown in Figure \ref{fig19}(7), while presenting the argument through narration that global adoption of meat-based diets would exceed the planet's available land resources, thereby encouraging viewers to think about animal conservation and sustainable living. }\end{quote}The accessibility and everyday nature of the featured content lower the threshold for engagement, effectively encouraging viewers to replicate these environmentally conscious actions in their own lives. Additionally, some videos cite speeches from leaders. \begin{quote}
    \textit{As shown in Figure \ref{fig19}(8), proactive measures not only create direct impact but also signal to others that meaningful engagement is attainable, inspiring participation through the power of example.}\end{quote} Creators present comprehensive information in an attempt to enhance its credibility, allowing viewers to intuitively experience the connection between individual actions and environmental conservation. 

\subsubsection{Objective Information Delivery Strategies in News \textcolor{blue}{and} Information Videos}\

In \textit{News and Information} videos, the focus on objective explanation and analysis results in a neutral emotional tone throughout the content. The information delivered in these videos can be categorized into three types: authority, relatability, and social identity. Various video segments play a crucial role in enhancing credibility, relevance, and practical guidance.

\begin{figure}[t]
    \centering
    \includegraphics[width=1\linewidth]{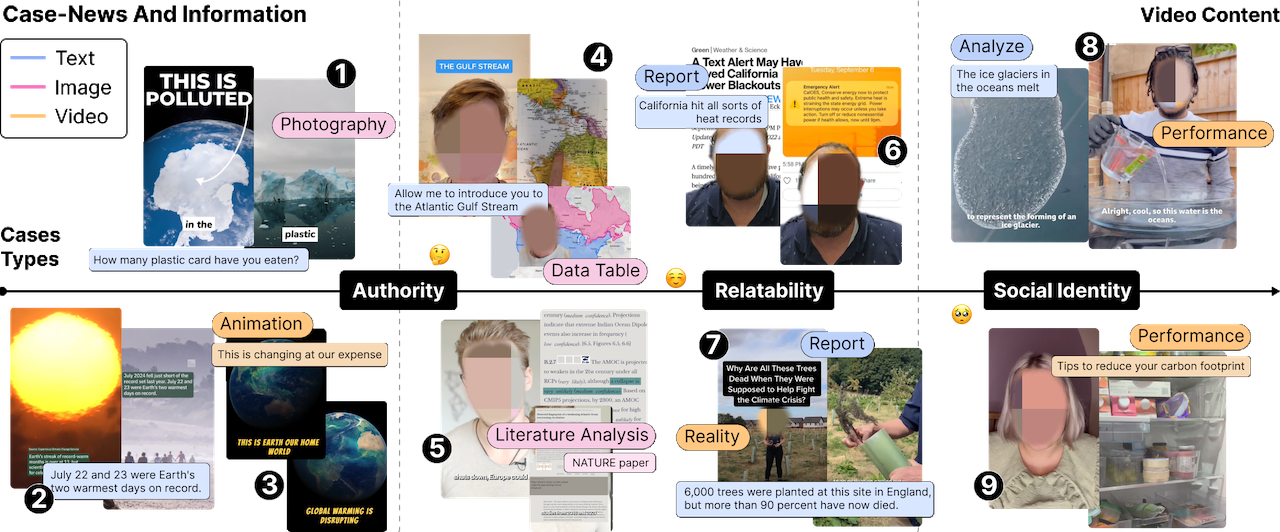}
    \caption{The trend of persuasion types in \textit{News and Information} and cases of video content}
    \label{fig21}
\end{figure}

Videos in this category attempt to convey authoritative information by quoting scientists or news anchors and presenting complex data charts to enhance the content's persuasiveness. \begin{quote}
    \textit{Some videos take viewers to remote locations, presenting evidence of plastic pollution in the world's most isolated places, highlighting the global scope and severity of the issue (Figure \ref{fig21}(1)). To further reinforce authority, certain official videos present real-time data (Figure \ref{fig21}(2)), such as extreme heat conditions, including reference to a specific period recently recorded as historically the warmest.}\end{quote} Presenting data attempts to strengthen the message and demonstrates the urgency of climate change. Data is often displayed dynamically through animated charts, accompanied by synchronized narration and subtitles to enhance the impact. By animating static data, creators adapt scientific rigor to the platform's fast-paced attention economy, ensuring that critical environmental metrics can compete with purely entertaining content.

Additionally, some videos use 3D models of the Earth to illustrate the progression of climate change (Figure \ref{fig21}(3)). Others employ maps and geography books to explain regional climate differences (Figure \ref{fig21}(4)), providing causal explanations such as the impact of ice sheet melt under conditions of an increase in global temperature. Some videos reference academic studies, analyzing research from journals like \textit{Nature} that examines phenomena such as the \textit{Atlantic Meridional Overturning Circulation (AMOC)} and its implications (Figure \ref{fig21}(5)). These predictions engage audiences with warnings that frame the timeline of potential outcomes within a span ranging from the near future to the end of the century.

\setlength{\textfloatsep}{5pt}

Relatability-based information connects with audiences through familiar situations. This highlights TikTok's unique strength in leveraging first-person, lived-experience narratives. For instance, \begin{quote}
    \textit{One video begins with a textual introduction that references a recent record-breaking heatwave affecting a coastal region, and proceeds to detail the consequential infrastructural impacts, such as widespread power disruptions (Figure \ref{fig21}(6)).}
\end{quote} \begin{quote}
    \textit{Another video takes viewers to an abandoned farm, showing the ecological decline from a first-person perspective (Figure \ref{fig21}(7)), while critiquing the implementation gap of a local afforestation initiative by pointing out the high mortality rate among newly planted trees intended for climate mitigation.}
\end{quote}This use of real-life examples makes the impacts of climate change more tangible, while creators leverage socially framed messages to encourage audience participation in environmental actions for social recognition. For example, \begin{quote}
    \textit{One video features a room tour where the creator shares eco-friendly habits (Figure \ref{fig21}(9)), such as reusing water for household plants to minimize waste.}
\end{quote} Another video demonstrates a climate science experiment. \begin{quote}\textit{As shown in \textcolor{blue}{Figure} \ref{fig21}(8), sodium acetate is used as a material to visually represent the physical process of glacier formation.}
\end{quote} Creators attempt to enhance understanding through interactive and participatory approaches, motivating viewers to apply the knowledge gained to their daily lives.

\subsection{Viewer Responses to Climate Change}
\label{sec:4.3}

\begin{figure}[t]
    \centering
    \includegraphics[width=1\linewidth]{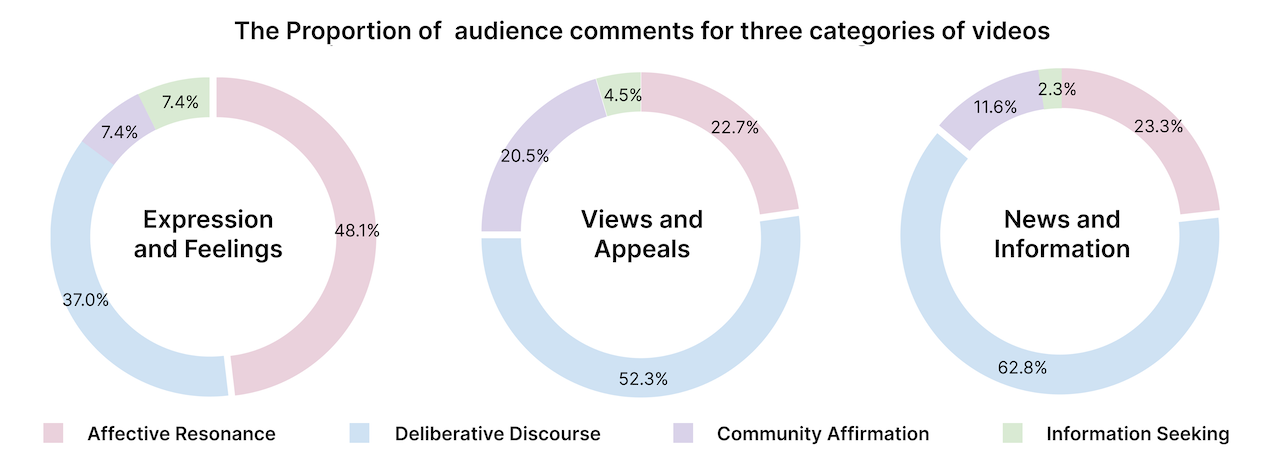}
    \caption{Proportion of audience comments for three types of videos}
    \label{fig22}
\end{figure}

We identified four distinct comment types and analyzed their distribution across video categories (Figure \ref{fig22}). ``Affective Resonance'' captures emotional reactions that interact with the video's tone, including direct expressions, deep resonance, and humor. The ``Deliberative Discourse'' involves critical thinking, debate, or persuading others, where audiences offer counterarguments, defend their positions, or attempt to challenge or support certain viewpoints, thereby fostering reflection and discussion. ``Community Affirmation'' provides positive feedback, motivation, or support, encouraging creators or other audiences. These comments often center around the topic discussed in the video. ``Information Seeking'' reflects audience interest in deepening their understanding or connecting with the content. This section analyzes the response trends and interactions between audiences, creators, and content across different video types.

\setlength{\textfloatsep}{5pt}

\begin{figure}[t]
    \centering
    \includegraphics[width=1\linewidth]{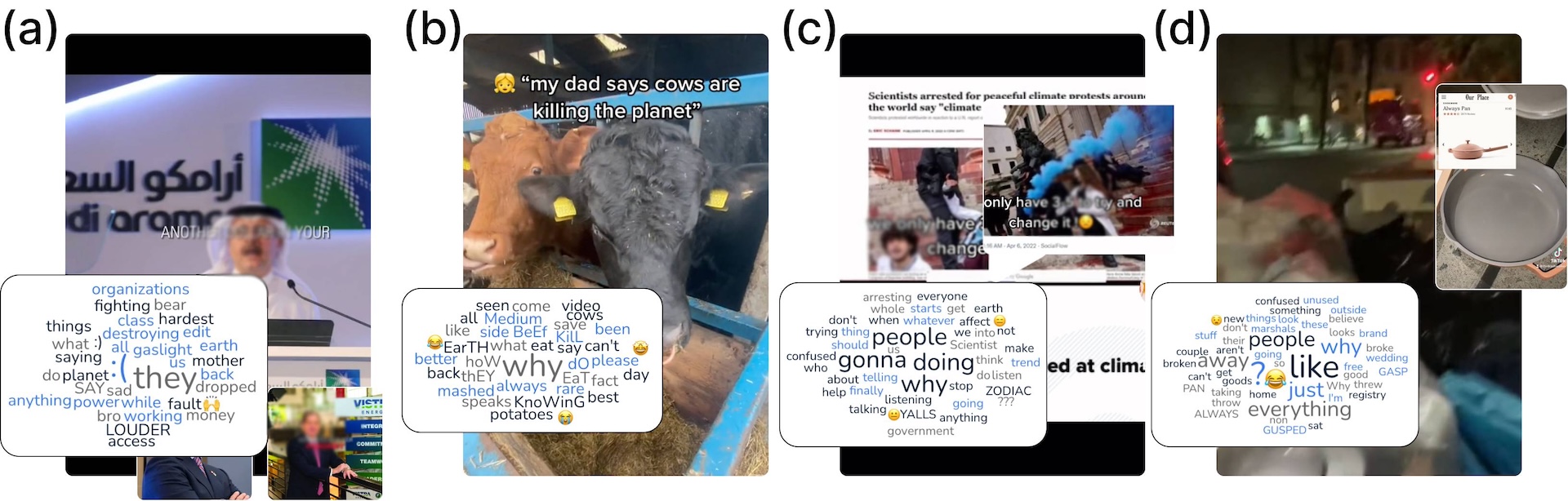}
    \caption{Video cases and comment keywords related to \textit{Expression and Feelings}}
    \label{fig23}
\end{figure}

\subsubsection{Comment Interactions in Expression \textcolor{blue}{and} Feelings Videos}\

In analyzing the three video categories, we found that viewer reactions exhibited significant diversity. \textit{Expression and Feelings} videos received the highest average number of comments, with an average of 1,236 comments per video. Of the comments analyzed, 48.1\% fell into ``Affective Resonance'' category; this high proportion reflects how the audiences \textcolor{blue}{utilized} the comment section as an outlet for venting. For example, \begin{quote}
    \textit{In a video featuring a montage of celebrities' portraits followed by a reversal highlighting current contradictions, one commenter noted the gaslighting of the working class by those in power, a sentiment echoed by others through crying emojis (Figure \ref{fig23}(a)).}\end{quote}
The use of emotionally charged words like \textit{sad} and \textit{crying} reflects the emotional connection with and resonance to the video content, while also serving as a means of expressing social identity through comments. This sense of identification is also demonstrated by how audiences are adept at responding based on the tone of the video. For example, viewers often mirror the creator's humor and tone. \begin{quote}
    \textit{In response to a video that mocked a father for blaming cows as a primary cause of ecological harm, a user humorously offers a justification for meat consumption as a mock solution to the ecological issue, while another expresses an eager desire to save the video to sustain the creator's humor (Figure \ref{fig23}(b)).}\end{quote}
Such playful interactions allow audiences to engage by using humor to alleviate the moral pressure associated with climate issues, thereby transforming serious topics into a form of social currency that fosters participation and deepens the discourse on climate change.

\begin{figure}[t]
    \centering
    \includegraphics[width=1\linewidth]{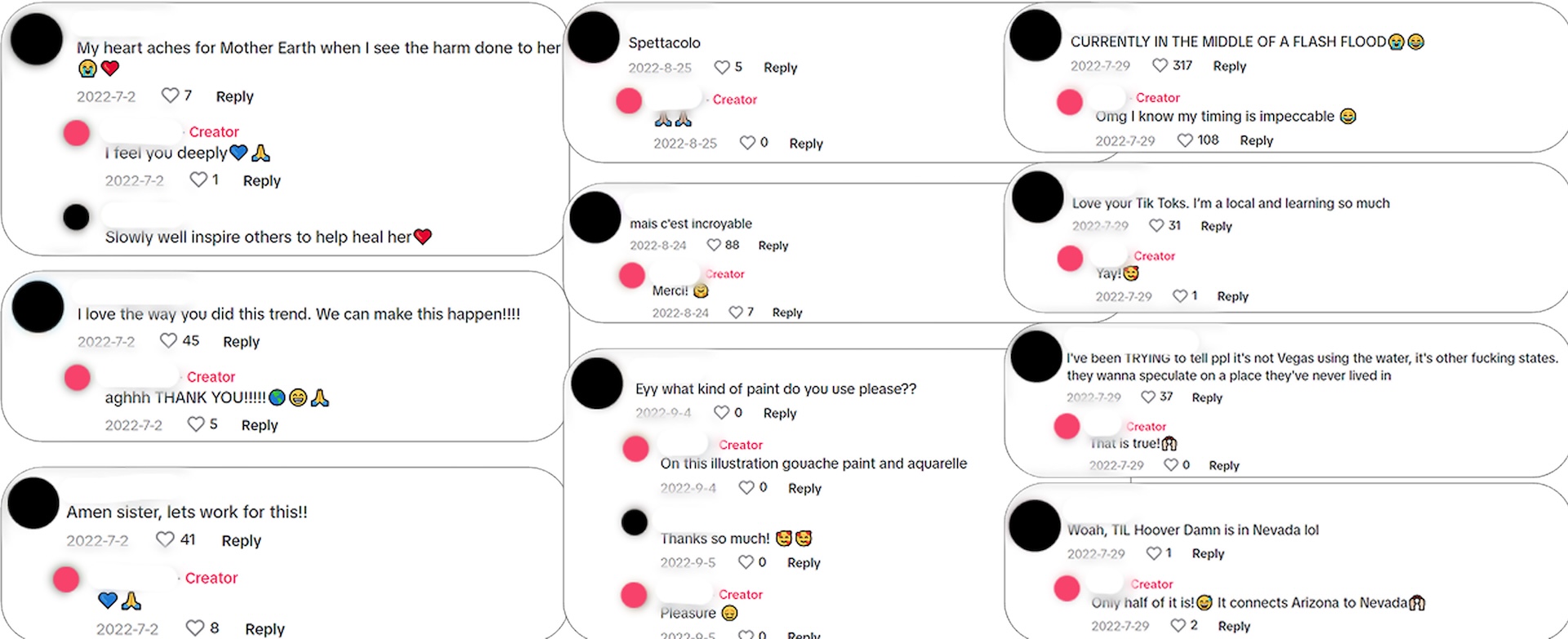}
    \caption{Interactions in the comments}
    \label{fig99}
\end{figure}

It is noteworthy that not all videos using emotion-driven strategies elicit resonance. An imbalance between emotional intensity and cognitive logic can instead prompt criticism. For example, in a video featuring an extreme doomsday narrative accompanied by news screenshots of climate activists being arrested, the comment section reflected widespread confusion.\begin{quote}
    \textit{Many users questioned the creator's intent and insisted that scientific warnings should be prioritized (Figure \ref{fig23}(c)).}\end{quote}
This reveals that high-pressure emotions may trigger defensive avoidance among viewers, which ultimately weakens the effectiveness of crisis communication. Furthermore, due to the diversity of audience values, even well-meaning actions might be criticized if they do not meet certain moral or behavioral standards. For instance,\begin{quote}
    \textit{A video sharing the experience of recycling furniture from a dump to encourage upcycling led to user confusion over the logistics and condition of the waste, questioned why such usable items would be abandoned rather than donated, casting doubt on the credibility of the upcycling process (Figure \ref{fig23}(d)).}\end{quote} 
While lifestyle-oriented content is easy to grasp, it can easily shift the viewers' focus from the environmental concept itself to a logical scrutiny of the video's narrative. Once the narrative presented deviates from audience common-sense standards, the original emotional appeal can transform into skepticism or even negative resistance.

In conclusion, \textit{Expression and Feelings} videos place more emphasis on emotional interaction. Creator interaction strategies tend to be emotionally engaging, often using emojis, exclamation marks, and emotional adjectives to express agreement and further resonate with the audience (Figure \ref{fig99}). This interaction model transforms climate change from a grand scientific proposition into a set of individually perceptible emotional experiences, thereby enhancing the public visibility of the issue.

\subsubsection{Comment Interactions in Views and Appeals Videos}\

\begin{figure}[t]
    \centering
    \includegraphics[width=1\linewidth]{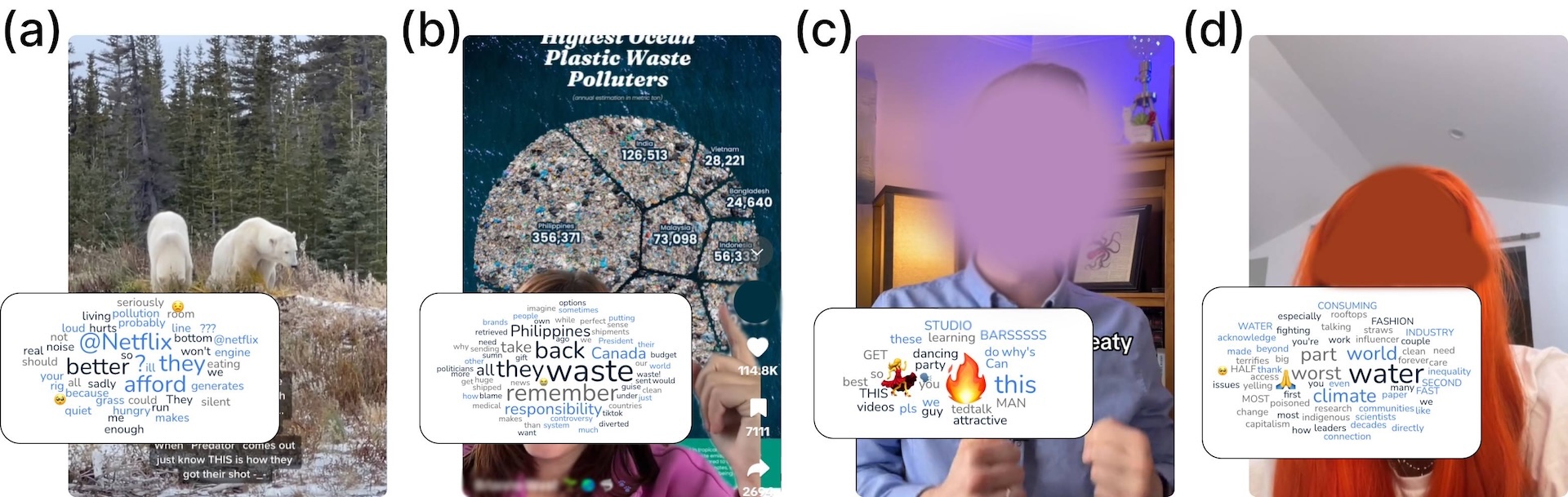}
    \caption{Video cases and comment keywords related to \textit{Views and Appeals}}
    \label{fig24}
\end{figure}

\textit{Views and Appeals} videos have an average of 1,208 comments per video, with 52.3\% of the analyzed comments categorized as ``Deliberative Discourse,'' 22.7\% as ``Affective Resonance,'' and the highest proportion of comments from all three video types being ``Community Affirmation.'' These data indicate that this type of video not only triggers intense content discussions but also elicits emotional responses and support from the audience. For example, in a video discussing the impact of noise pollution on polar bears, shared from a Netflix climate documentary, audiences engage in both emotional expression and rational analysis. \begin{quote}
    \textit{Comments regarding the prioritization of corporate profits over ethical responsibilities, alongside observations of extreme deprivation, reflect a combination of critical public opinion and deep-seated emotions such as anger and sympathy (Figure \ref{fig24}(a)). }\end{quote}

The audience's emotional fluctuations regarding climate issues immediately trigger an impulse to seek causes and assign responsibility. The narrative style of these videos uses engaging stories to convey their perspectives, guiding viewers to find logical anchors for their emotions, gaining both emotional and intellectual approval from the audience. \textcolor{red}{\sout{, as reflected in comments that}}\textcolor{blue}{For instance, comments often} question the efficacy of international politicians while suggesting that TikTok creators often provide more relatable or sensible insights into global issues (Figure \ref{fig24}(b)).

An interesting finding is that these videos also inspire audiences to thank the creators personally. For instance, one creator humorously explains the dangers of high temperatures and humidity to health while humming along. Compared to large, cold, official institutions carrying political labels, the authentic persona displayed by such creators lowers the defense threshold for information. This transforms climate knowledge from a condescending lecture into trustworthy sharing of information from a peer. Audiences responded with comments, \begin{quote}
    \textit{Celebrating the content as a unique synthesis of education and entertainment, with many advocating for the creator to receive professional production support (Figure \ref{fig24}(c)).}\end{quote} 
Additionally, another creator called for climate action during a conversation with friends, \begin{quote}
    \textit{Leading to feedback from climate researchers who praised the creator for filling a void in influencer-led environmental communication. This interaction underscores the perceived importance of mainstream voices in validating scientific concerns (Figure \ref{fig24}(d)).} \end{quote} 
This phenomenon not only demonstrates the potential of personalized storytelling and entertaining delivery to enhance audience engagement and emotional connection, but also shows how the public shifts its focus from macro issues to personal expressions, making the topic more tangible. Furthermore, this act of thanking the creator is, in essence, an acknowledgment of the authenticity and personification expressed within climate discourse.

\begin{figure}[t]
    \centering
    \includegraphics[width=1\linewidth]{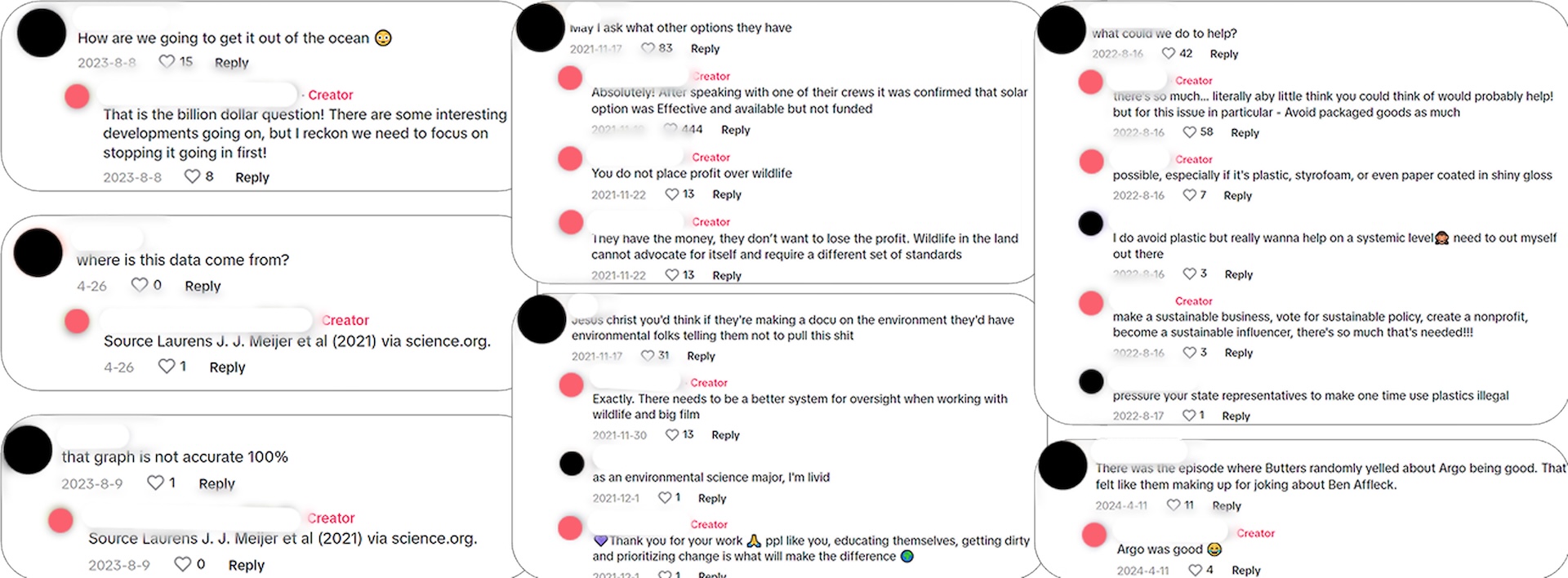}
    \caption{Interactions in the comments}
    \label{fig98}
\end{figure}

In summary, \textit{Views and Appeals} videos balance emotional expression and persuasion to stimulate audiences to express their personal viewpoints. Creators deepen the core message of the video by responding to viewers' questions or further analyzing comments in the comment section, sometimes attracting more comments and sparking more in-depth discussions. This interactive approach makes the comment section more discussion-oriented, with viewers asking for more details or seeking advice. Creators often use call-to-action language, such as \textit{take action} and \textit{we should}, to motivate audiences to participate in the discussion or take real-world actions (as shown in Figure \ref{fig98}). Through the creators' guidance, what audiences engage in within the comment section is not merely a clash of opinions, but an exercise to simulate and affirm future actions.

\subsubsection{Comment Interactions in News \textcolor{blue}{and} Information Videos}\

\begin{figure}[t]
    \centering
    \includegraphics[width=0.8\linewidth]{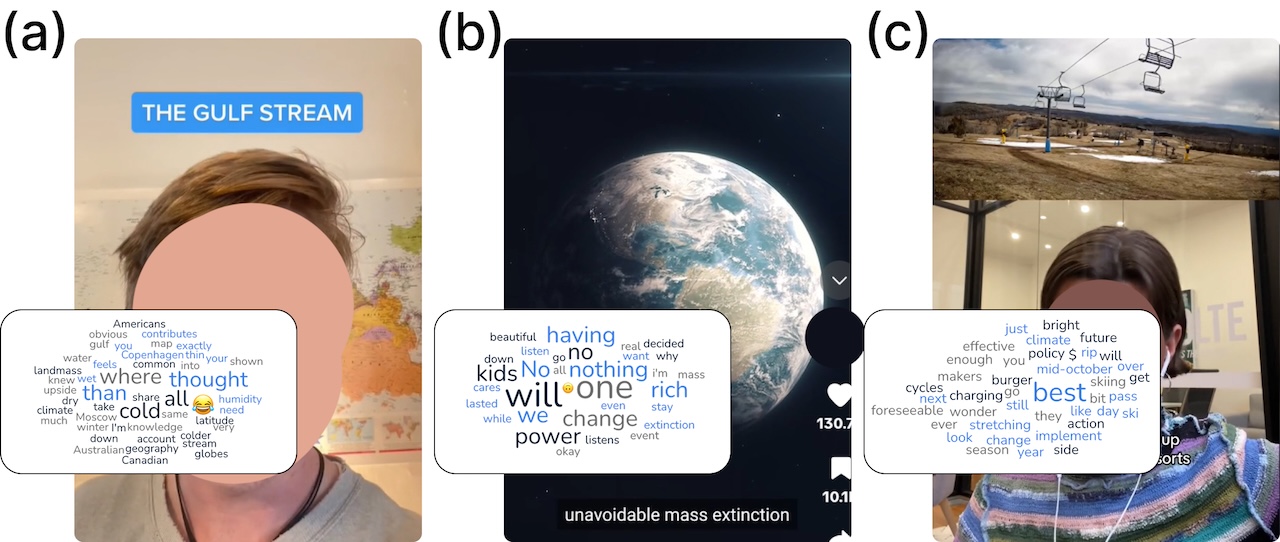}
    \caption{Video cases and comment keywords related to \textit{News and Information}}
    \label{fig25}
\end{figure}

In contrast, \textit{News and Information} videos attract comments that tend toward rational discussions. These videos receive an average of 1,072 comments, with 62.8\% of the analyzed comments categorized as ``Deliberative Discourse.'' For instance, in a video explaining why the UK is colder than the US despite being on the same latitude, \begin{quote}
    \textit{The comments reflect a collaborative exchange of geographical knowledge, with viewers highlighting how Europe's coastal shape influences its weather patterns. Others pointed to the importance of atmospheric moisture, arguing that high humidity levels in regions like Ireland make winter temperatures feel significantly more severe than the data might suggest (Figure \ref{fig25}(a)).}\end{quote} 
These discussions revolve around geographical knowledge without emotional language, indicating that audiences engaged with these videos by analyzing the facts rather than expressing emotions. This tendency may stem from audiences' desire for knowledge as they seek deeper understanding through discussion. Viewers shifted the focus of their interaction toward addressing knowledge gaps and fact-checking; at this point, the comment section functions more like a collaborative space for online learning.

Interestingly, even after receiving factual information, audiences sometimes express emotions to intensify their sense of conflict and place blame. For instance, a video shared predictions from Europe's most powerful supercomputer.
\begin{quote}
    \textit{Commenters highlighted a deep-seated pessimism, reflecting a belief that institutional inertia would prevent any meaningful change despite the alarming scientific predictions (Figure \ref{fig25}(b)).}\end{quote} 
Another video shared data on Australia's recent poor skiing seasons, and instead of further discussing the data, comments pointed directly at the government: 
\begin{quote}
    \textit{\textit{Comments regarding whether policymakers' personal interest in skiing might finally spur climate action, alongside complaints about inflated prices at resorts, suggest a shift toward blame as the environmental crisis becomes undeniable} (Figure \ref{fig25}(c)).}\end{quote} 
When the problem becomes impossible to ignore (especially for ski enthusiasts), viewers look for someone to blame. This behavior serves to alleviate the psychological pressure caused by receiving negative forecasts, transforming the sense of helplessness regarding climate decline into accountability for institutional authorities. This blame-oriented interaction places specific demands on the feedback mechanisms of content publishers.

\begin{figure}[t]
    \centering
    \includegraphics[width=1\linewidth]{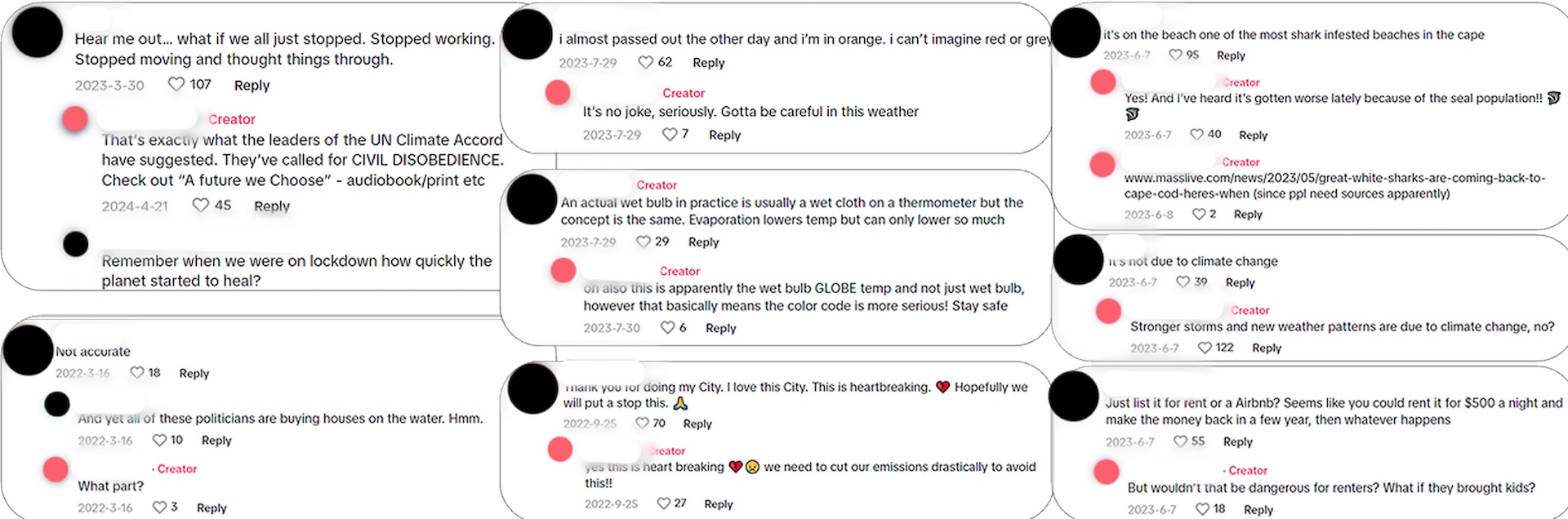}
    \caption{Interactions in the comments}
    \label{fig97}
\end{figure}

Regarding interactive content, \textit{News and Information} videos present an authoritative interaction paradox. While official videos have high engagement, institutional accounts generally remain silent and do not engage with audiences. This silence maintains distance from the audience to safeguard official authority. Individual creators frequently respond to questions, using discussions or counter-questions to strengthen their stance. This closely relates to the educational nature of \textit{News and Information} videos (as shown in Figure \ref{fig97}). Direct replies from creators provide a pathway for knowledge sinking. This problem-based interaction maintains the educational function and soothes climate-related stress, filling the interactive void left by official institutions.

\section{Discussion}\label{sec:Discussion}
\subsection{Strategic Expressions and Interactions of Content Creators in TikTok Climate Communication: Balancing Media Incentives and Information Accuracy}\

This study is based on the framework of climate communication theory. It examines the dissemination patterns of climate-related videos on TikTok. Specifically, it explores how creators' strategies and audience interactions shape four distinct communication types. These categories reveal how creators simplify complex information through strategic formats to foster discussion. This ecosystem broadens public participation, while also revealing that the effectiveness of communication largely depends on individuals' evaluations of the content and their willingness to further disseminate it~\cite{Moser}. At the same time, opaque algorithmic mechanisms incentivize creators to iteratively learn platform preferences to maximize visibility~\cite{bucher2019algorithmic, karizat2021algorithmic}. The specific strategies we observed thus reflect adaptive efforts to balance engagement with informative value. These findings provide valuable practical insights for building more effective climate communication mechanisms on social media in the future.

\subsubsection{Portraying Climate Issues Through Emotions}\

\textit{Expression and Feelings} videos use emotional storytelling, personalized delivery, and sentimental editing to strengthen the affective dimension of climate communication. Compared to directly criticizing social issues, policies, or power institutions, which might lead to controversy, platform bans, or audience resistance, emotional and humorous packaging delivers critical information in a seemingly harmless manner. For example, exaggerated emotions such as anger, helplessness, and absurdity are expressed through funny expressions, catchy tones, rap, or exaggerated acts of anger (Figure~\ref{fig17}). This method both conveys an attitude and preserves social engagement with ambiguity. 

Although such content is often criticized as information fragmentation or low-quality information~\cite{castaldo2022junk}, this format is effective and necessary within current platform ecosystems. On one hand, as serious news and political content often struggle to go viral~\cite{hagar2025algorithmic}, emotional expression is more likely to be favored by algorithms on short-video platforms~\cite{shutsko2020user}. On the other hand, absurdity and emotional tension capture more attention, triggering broader algorithmic distribution of climate issues~\cite{siles2024learning}.

In our case studies, many exaggerated negative emotional performances are found, often with high emotional tension and rapid shifts in video pacing, evoking a \textit{laughing while cursing} or \textit{cursing while sympathizing} response from the viewers. Other strategies range from gentle narratives juxtaposing scenery with tragedy to artistic metaphors, creating a multi-layered emotional impact. From a communication psychology perspective, affective activation typically precedes cognitive processing~\cite{lodge2005automaticity}, allowing high-arousal emotional content to rapidly break through informational noise and capture audience attention. In this process, strong personal stances create memory points while aligning with platform incentive logic. Since TikTok utilizes computer vision and natural language processing to categorize visual and audio components~\cite{siles2024learning}, a stable style, such as a creator who consistently wears the same costume, establishes a recognizable personified pattern, securing consistent traffic recommendations and follower interaction.

Research shows that both positive and negative emotions can motivate sustainable behaviors~\cite{Fritsche, cox2013environmental}. Positive emotions stimulate the willingness to take action, while negative emotions like guilt are amplified in social media's collective atmosphere, increasing sharing intent~\cite{Rees, Vlasceanu}. Although direct behavior translation remains under-studied, when these emotions are combined with humor, exaggeration, and plot twists, they help enhance information retention and deepen the audience's impression of climate issues~\cite{Brosch, Liang}, thereby contributing to improved video metrics.

This emotional pattern helps creators gain initial traffic and stimulate deep viewer involvement, leading to more comments. Our research found that content creators tend to use emotionally charged language (e.g., sadness, anger), emojis, and emphatic expressions to guide viewers toward emotional resonance and value alignment. Besides, viewer comments serve as both emotional responses and secondary creations, ranging from empathetic reinterpretation to the negotiation of opposing viewpoints. Viewers' attitudes are influenced by the evoked emotions and the characteristics attributed to them. This guides thinking toward negative or positive traits~\cite{zerback2024role}, reflecting the potential of this strategy to shape public attitudes toward climate issues.

Comment analysis shows that ``Affective Resonance'' accounts for nearly half of these video responses. These comments include direct emotional expressions, personal reflections, and discussions with a sarcastic tone. This demonstrates that content echoing viewers' climate-related feelings (such as helplessness, anger, or anxiety) triggers a strong drive for self-expression. This ``I feel the same way'' resonance motivates them to transform their feelings into viral, replicable memes~\cite{martin2019tiktok}. In this process, viewers find community and validate their emotions. Faced with the complex nature of climate change, expressing emotions becomes a low-cost way of emotional release. Whether supportive, critical, or skeptical, these interactions generate the high-frequency data that incentivizes algorithms to boost traffic~\cite{maris2025taking}. This process turns engagement incentives into public visibility for climate issues.

Therefore, \textit{Expression and Feelings} videos play a dual role in climate communication. On one hand, they enhance emotional appeal through affective information presentation; on the other hand, they trigger diverse interactive behaviors, which boosts visibility under algorithmic incentives, ultimately transforming emotional engagement into broad public participation.

\subsubsection{Self-Projection in Issue Response}\

\textit{Views and Appeals} videos adopt a value-driven discursive framework to modularize grand and abstract climate issues, such as biodiversity loss~\cite{Delmotte, Hong, Baird}, the fossil fuel crisis~\cite{David, Abbass}, and electronic waste pollution~\cite{Devin}. Creators typically avoid complex academic jargon, opting instead\textcolor{red}{\sout{ instead}} for a strong self-disclosure style that places the individual at the core of the narrative. Their videos typically center on them speaking in a close-up shot, through which they share third-party experiences (such as reported events, media coverage, or social problems) or specific occurrences (such as wildfires, family indifference toward climate issues, or personal environmental practices) to visualize and personify the climate agenda (Figure~\ref{fig19}). This approach possesses dual efficacy: specific events elicit public empathy, as people are more concerned with how climate change affects daily life rather than the complexity of the problem itself~\cite{Susie, Elsayed-Ali}; furthermore, modularized content facilitates precise algorithmic distribution.

These videos also exhibit strong mobilizing characteristics, with creators offering small and specific action suggestions through snippets of daily life (such as deleting spam emails or rejecting plastic straws) after expressing their views, providing clear paths for participation. This not only enhances the creators' sense of meaning on a behavioral level~\cite{brosch2021leveraging}, but also stimulates positive cognition and emotional responses in the audience~\cite{Venhoeven}. This transition from \textit{talking} to \textit{doing} in the video content aligns with Benke's \textit{positive experience-behavior motivation model}~\cite{Benke}, highlighting its function in motivating others. More importantly, these specific calls to action induce user-generated content. Viewers provide feedback on their actions in the comments, generating high-frequency interaction data that triggers platform incentives, sustaining the topic's visibility~\cite{maris2025taking}.

These videos demonstrate a bidirectional mechanism of expression and negotiation between creators and viewers. On the one hand, creators construct an inspiring and directive communication style through narrative storytelling, cautionary slogans, and emotionally charged appeals. On the other hand, viewers actively participate in issue discussions through responses such as ``Deliberative Discourse'' or ``Community Affirmation'' in the comments, feeling that ``I also have something to say.'' Faced with specific events, viewers show a willingness to become co-creators of viewpoints~\cite{baden2017conceptualizing}, blending emotional resonance with critical thinking. Comment sections show a dialogic structure: creators answer questions and extend discussions, while viewers debate or seek guidance. For example, comments like ``How can I help?'' spark interactions that make actions more concrete and relatable, while also attracting wider attention through likes and further comments. This support shows viewers value the issue, motivating them to stay informed~\cite{boningerleandre} and reflect deeply~\cite{petty1990involvement}.

When viewers discover that the creator expresses political views, moral beliefs, or lifestyles that align with their own, they feel a sense of ``we are the same type of people.'' Expressing gratitude or support becomes a low-cost, high-emotion return action. When direct action is impossible, online support serves as an alternative form of participation. This structure turns the comment section into an extended space for the public to collaboratively construct climate understanding and action frameworks, enhancing both the intensity of the topic's discussion and the interactivity of the communication, helping the platform algorithm recommend the video. Therefore, \textit{Views and Appeals} videos contribute to shaping climate attitudes through content creation while simultaneously constructing a participatory and engaging model of climate communication through discursive negotiation and emotional resonance between creators and viewers.

\subsubsection{Knowledge Reshaping and Negotiation under Media Incentives}\

\textit{News and Information} videos play a knowledge-centered role in climate communication, yet they are not overly serious. While these videos are typically released by knowledge-based influencers or official media accounts, \textcolor{red}{\sout{but }}TikTok gives creators greater discursive power. Faced with climate change, some creators choose not to use emotional outbursts or personal advocacy. Instead, they use educational videos to help viewers understand facts and make informed decisions. To manage complexity, creators attempt to package scientific knowledge in \textcolor{blue}{an} accessible, engaging, and shareable way. 

This type of content emphasizes logical clarity and factual accuracy. Even when discussing concrete topics such as forests, children, or urban heatwaves, these videos remain analytical. They use fun experiments (in their backyard), lively animations, or facial expressions to make these videos knowledgeable yet not serious. This shift not only aligns with young audiences' social media habits~\cite{hendrickx2025normal} but also transforms dry scientific reports into a highly attractive digital space through visualization and emotional narrative~\cite{wahl2020emotional}. At the same time, these videos demand a higher level of cognitive engagement from the audience, but they encourage curious viewers to rewatch and deepen their understanding through comments, breaking the bias that short videos cannot facilitate deep understanding~\cite{Yunpeng}, showcasing the potential for deep communication within short content. Moreover, through the platform's recommendation algorithms, repeated viewing further reinforces the exposure to such content, gradually deepening viewer engagement and comprehension over time. Repeated viewing reinforces exposure and deepens viewers' commitment.

Contrary to the perception that such videos are mostly produced by knowledge influencers or official media accounts, TikTok creators often adopt rich explanatory language, data visualizations, and citations from authoritative research (e.g., through the repeated use of charts, news clips, or literature screenshots) to enhance credibility. This finding further supports research on climate communication on social platforms~\cite{fernandez2016talking, falkenberg2022growing}, suggesting that climate communication on social media relies on visual symbols and personalized narratives for appeal. This shift fits digital consumption habits rather than threatening journalistic norms, creating a news space that blends emotion with rationality~\cite{wahl2020emotional}.

Interactions vary by creator identity. Some creators often answer inquiries, forming a chain of knowledge supplementation and extended dissemination. In contrast, some institutional accounts, despite their authoritative standing, tend to maintain interactive silence as a strategic choice, resulting in what can be described as authority-driven detachment. This silence preserves authority but disadvantages the content in the social media ecosystem. Low interaction reduces algorithmic weight, limiting the video's reach. This confirms that interaction is essential for visibility on social platforms~\cite{maris2025taking}.

At the viewer level, comments are characterized by ``Deliberative Discourse,'' where users typically supplement, discuss, or challenge the facts and logic, forming a knowledge-negotiation interactions. To overcome information overload, viewers seek simplified, accurate interpretations, prompting creators to provide further explanations or references. These interactions expand video reach and shift viewers from passive receivers to active cognitive collaborators. Through comments, viewers, facing the sense of helplessness when confronted with complex issues, gain a small yet real sense of agency by saying, ``I understand it too.'' This reflects the viewer turn in journalism research, where individuals or collectives exert power over issues through intentional or accidental actions~\cite{hendrickx2023power}. It is also worth noting that although this type of video generally exhibits a lower level of emotional expression, under specific topics—such as biodiversity loss and extreme climate predictions—viewers may reinforce their stances through emotional comments, expressing helplessness, criticizing governments, or questioning slow responses to climate issues. This reveals a public craving for authoritative responses to the climate crisis. It also underscores the importance for official communication channels to rebuild dialogue mechanisms within digital spaces.

In summary, \textit{News and Information} videos, through rational expression and knowledge negotiation mechanisms, present complex climate knowledge in a de-authoritized way that is accessible to the public. These videos often piece together fragmented information to create an expression with the characteristics of scientific interpretation, establishing credibility within an informal context. In this process, content creators act as knowledge providers and framers, while viewers engage through supplementation, correction, and in-depth discussion, helping to collectively advance the knowledge-based and socialized understanding of climate issues. This signals a trend of news transformation on social media: news is no longer a closed finished product but a dynamic space where creators and audiences co-evolve through algorithmic mediation.

\subsubsection{Trend Hijacking}\

Although \textcolor{blue}{\textit{Trend Hijacking}} videos are not directly related to climate issues, they utilize climate-related hashtags to attract traffic, forming an informal mode of participation in climate communication. While this type of atypical participation does not directly trigger action, its role in boosting topic visibility and activating platform algorithms should not be underestimated. As research has pointed out, public attention to climate issues does not always manifest through deep cognitive engagement; increasing topic visibility itself is one of the key goals of climate communication~\cite{Schafer}. On the other hand, \textcolor{blue}{\textit{Trend Hijacking}} is not just a strategy to capitalize on trends; it also reflects the dynamic nature of user participation and information dissemination on the TikTok platform. By participating in popular trends or using trending hashtags, creators break traditional communication paths and leverage the platform's algorithmic recommendation system to spread content to a broader audience. User interactions (such as likes, comments, and shares) drive the dissemination of information, which spreads rapidly across different nodes~\cite{pathak2023understanding}. The connections between these nodes are diverse and interwoven, leading to varying pathways and speeds of information transmission. Unlike traditional linear communication methods, this type of dissemination is influenced by multiple factors, such as user interaction, platform algorithms, and hashtags, allowing information to circulate across multiple directions, platforms, and social circles, ultimately expanding its reach and influence.

\subsubsection{The Strategic Trade-off: Balancing Media Engagement and Scientific Accuracy}\

This study reveals that climate communication on TikTok is not direct fact transmission, but a product of tension between \textit{media incentives }and \textit{information accuracy incentives}. These interactions create video categories that reflect the trade-off between attracting audiences and delivering facts. \textit{Expression and Feelings} and \textit{Trend Hijacking} prioritize engagement. Using emotional narratives, satire, or trending topics, creators turn dry science into high-arousal media. Despite risks of fragmentation and controversy, this approach helps the issue break out of the circle and overcomes audience isolation. \textit{Views and Appeals} occupies the middle ground, using first-person narratives to anchor macro-issues. This format bridges emotion and logic by combining scientific discourse with personified appeal. \textit{News and Information} videos represent the highest effort toward accuracy. These creators use experiments and animations to make science visually attractive without sacrificing core facts. The trade-off enhances visibility by wrapping scientific facts in an emotional shell.

\subsection{Insights for Climate Communication Targeting Content Creators and \textcolor{blue}{Viewers}}\

The rise of social media has reshaped the pathways of climate communication, providing the public with important platforms for expressing opinions, sharing experiences, and engaging in collaborative participation~\cite{Silvia}. Through these platforms, the public can achieve emotional resonance and experience exchange with others, rather than relying solely on governmental or expert solutions. In this public (content creators)–public (viewers) communication environment, video content serves not only as an information carrier but also as a catalyst for opinion exchange and cognitive negotiation. Content creators convey attitudes and emotions through their videos, while viewers respond through comments, sharing, and content imitation, forming a collaborative and emotionally rich communication practice. By explaining how public engagement and platform incentives drive communication, our findings further expand on the characteristics of TikTok as a communication channel within climate communication theory~\cite{mavrodieva2019role}. Moreover, it provides a multidimensional perspective for understanding climate agency within digital spaces.

\subsubsection{Creator Strategies: Personal Narratives, Humor, and Safety}\

For climate creators and educators, diverse narratives lower the understanding threshold and provide guidelines for generating meaningful audience feedback. Social media creators excel at building personal narratives to make their concerns relatable and attract viewers~\cite{zhao_iwas_2026,zhang_laughing_2026}. In the context of personal narratives, video creation or comment interaction becomes an effective way to express concerns about climate issues. For high-controversy and polarizing topics (such as climate policy responsibility and government inaction), creators can adopt satire and facts (e.g., skits or role-playing) to indirectly guide viewpoints. Additionally, creators can further use rhythmic background music, absurd dialogues, and repeated keywords to highlight humor and exaggeration, employing soft expression strategies that convey stances while avoiding bans or confrontation.

Research has shown that humorous content often carries meaningful political work, including expressing opposition, political identity, and showcasing civic support~\cite{davis2018seriously}. This study confirms this finding and reveals how \textcolor{blue}{content creators} use political humor to express their personal views on climate issues. This is an effective safe expression strategy in complex environments. Meanwhile, such expressions emphasize imitativeness and emotional resonance, helping creators build personal styles and fan bases. Regardless of whether the emotions are positive or negative, emotional resonance is an indispensable factor~\cite{Kollig}. This also explains why many videos highlighting climate-related issues, such as floods, emphasize specific local regions, guiding residents to share real-time updates through comments and thereby stimulating deeper discussions through concrete scenarios. Moreover, content creators can also start from their own experiences, using small and specific life stories to draw attention, making climate issues more relatable to daily life. For example, linking temperature charts to personal heat discomfort helps viewers see the real impact of climate change. This emphasis on personal experience avoids the cognitive burden of complex scientific language or policy analysis when discussing climate issues. Instead of directly conveying viewpoints, it subtly guides viewers to form their perspectives through storytelling. 

For creators skilled in organizing information and possessing professional knowledge, they can build authority through informational videos. However, adding emotional expression to facts and policy analysis improves relatability. Using animations and analogies improves visual appeal and prevents content from appearing too official~\cite{schuster2024being}. Everyday metaphors and personal reactions (e.g., ``I was shocked'') help guide emotions and increase resonance. These findings further support climate communication theory, demonstrating that communication strategies can significantly influence viewers' attention to climate issues~\cite{weber2010shapes}. This explains why official accounts remain cautious to maintain a controlled discussion environment. This study extends the theory by revealing that communication strategies from a personal perspective—particularly through creators' personal narratives, humor, and emotional resonance—are more effective in uncovering the complex issues in climate communication and engaging viewers' attention.

\subsubsection{Fostering Collaborative Agency and Algorithmic Training}\

Climate communication theory emphasizes the importance of communication in climate messaging~\cite{moser2010communicating}. Therefore, climate communication on social media should better utilize the comment section to enhance interaction and collaboration between content creators and viewers. These behaviors are shaped by the platform's incentive mechanisms. TikTok's algorithm uses metrics such as likes, comments to determine video exposure~\cite{tiktok2020tiktok}. Thus, creators must both attract audiences and intentionally encourage engagement.

Specifically, creators can guide discussions by asking open-ended questions at the end of videos, such as ``What are your thoughts on this issue?'' or ``What actions do you think we can take?'' to encourage viewers to share personal insights and stimulate deeper conversations. Creators can also regularly respond to comments—especially meaningful ones—and pin them to increase visibility, fostering a greater sense of community engagement and belonging. Additionally, creators can launch interactive polls and challenge campaigns (e.g., ``Challenge yourself to go plastic-free for a day'') to encourage viewers to participate in concrete climate actions, thereby enhancing their sense of responsibility and involvement. Through these approaches, content creators can foster a collaborative, emotionally resonant, and action-oriented environment for climate communication, motivating the public to actively engage in discussions and real-world practices related to climate issues. In essence, this represents a strategic adaptation by content creators to platform incentives: by stimulating high interaction data to please the algorithm, they maximize the visibility and dissemination efficacy of climate-related content.

Viewers can actively engage in comment interactions by expressing views on climate, posing questions, sharing personal experiences, or critically reflecting on existing content to promote deeper discussions, much like in narrative content~\cite{cao_audience_2026}. For instance, viewers can share information about local climate conditions or offer insights into their own climate actions and experiences expands collective knowledge. Additionally, viewers can support calls to action by responding to challenges or polls initiated by content creators, further encouraging tangible actions. Through liking, sharing, commenting, participating in trending discussions, or sharing the outcomes of their climate actions, viewers not only help expand the reach of climate topics but also inject greater momentum into climate discourse on social media platforms. Furthermore, viewers can initiate non-instrumental topics (e.g., zodiac signs and color leanings) to bypass censorship and simulate deep engagement~\cite{maris2025taking}. These strategic threads maintain activity and influence how the algorithm allocates traffic.

Viewer interaction is both emotional expression and an active participation strategy. \textcolor{blue}{``Filter bubbles''\footnote{\textcolor{blue}{A state of intellectual isolation where personalized algorithms automatically screen out dissenting information, leaving users in an invisible, self-reinforcing web of their own interests.}} and ``echo chambers''\footnote{\textcolor{blue}{A social or digital space where specific beliefs are amplified and reinforced by repetitive communication within a like-minded group, effectively drowning out opposing viewpoints.}} are often \textcolor{red}{\sout{overestimated}} viewed with concern~\cite{jones2024can}, yet such algorithmically formed communities may also play a constructive role.} Audiences have agency: when users tired of repetitive content~\cite{siles2024learning} or accidentally encounter new information~\cite{hagar2025algorithmic}, they are likely to engage in conscious information processing~\cite{jones2024can}. This includes actively searching for new keywords or intentionally clicking on new topics to break through information cocoons. Every user interaction trains the algorithm in real-time~\cite{maris2025taking}. Through this strategic voting, users express stances and proactively reshape climate issue visibility in their information streams. Therefore, even if ``information bubbles'' exist to some extent, the lack of clear community boundaries on TikTok may actually play a positive role. By grouping users with shared identities, TikTok builds climate communities via algorithmic logic~\cite{maris2025taking}.

Consequently, TikTok's climate communities are fluid landscapes co-created by algorithms and interaction. When a video goes viral, a temporary discussion community forms rapidly in the comments. The community is a dynamic process driven by creators seeking exposure and viewers shaping their feeds. Climate communication should utilize these mechanisms, making interaction the core strategy to generate ripples within algorithmic waves.

\subsubsection{Leveraging Platform Features and Future Directions}\

TikTok's format shows how social media can broaden climate communication's reach. Content creators' diverse strategies attract repeat views, which the algorithm amplifies to increase issue visibility. Audience interaction further reinforce this amplification effect. Its concise content and viral spread potential make climate issues more accessible to the general public~\cite{Le}. Furthermore, TikTok's production assets, like music and stickers, significantly simplify the creation process. This encourages active replication over passive watching, turning science education into a viral, community activity. Users can join discussions simply by sharing analysis or editing video fragments with music. This ease of participation lowers creation barriers, allowing diverse voices to join the global climate conversation. Looking forward, climate communication should continue leveraging platform features to enhance awareness and impactful action. 

While the underlying logics of various short-video platforms, differ in terms of interaction weighting and distribution mechanisms~\cite{Wang2021Understanding}, strategies like emotional resonance, personification, and humor remain largely universal. These strategies not only work well with TikTok's algorithm but can also capture attention and engage users on other platforms.

Simultaneously, this research provides strategic insights for cross-media advocacy. Instagram can move beyond perfect aesthetics~\cite{sulik2014rethinkpink} by using raw, real-life footage to show the gravity of climate change. YouTube Shorts can act as a gateway, driving viewers from quick clips to in-depth educational content. Rednote can turn complex science into practical, everyday tips for its users. Social platforms enable image management in self-presentation~\cite{zhang_image_2025}. However, short videos cannot replace the role of professional news and long-form media. In-depth reporting and policy debates still require professional media. Therefore, future climate communication should leverage various platform strengths to create a holistic discussion space.

However, communication efficacy cannot rely solely on creator and audience agency. Short-video platform algorithms prioritize clicks over quality, often drowning serious issues in entertainment noise. This highlights the necessity for platforms to implement structural interventions~\cite{hagar2025algorithmic}. Examples include weighting authoritative sources or establishing dedicated news sections outside the standard feed.

\subsection{Limitations}
Multiple limitations may affect the comprehensiveness and depth of the research findings when researching climate change-related content on the TikTok platform.

\subsubsection{Ethics in Social Media Content Analysis} Despite data being public, users may not have consented to its use in academic research. The dataset's scale makes it difficult to gauge individual willingness to participate. Therefore, we must acknowledge that if an external party has sufficient intent, they may still be able to re-identify the original poster by cross-referencing the detailed quotes and descriptions presented in this paper with publicly available information. Addressing this requires ongoing dialogue in CSCW and HCI to develop context-sensitive ethical guidelines. For instance, social platforms could allow users to opt-in or out of academic research. We also urge future social media researchers to pay attention to user privacy issues and explore methods that can safely balance analytical depth with user protection.

\subsubsection{Platform Limitations} Researching solely on the TikTok platform may not provide a complete understanding of public perceptions regarding climate change. Although TikTok is particularly popular among younger audiences, its user demographic and content formats differ significantly from other social media platforms \textcolor{blue}{X} and Facebook. These platforms may offer different discussion environments and perspectives, which could lead to varying understandings and reactions to climate change. Moreover, we cannot accurately determine the specific composition of the public or viewers on social media, and how they encountered the analyzed videos.

\subsubsection{Language Barriers} As a global platform, TikTok features videos and comments in multiple languages. However, in this study, due to language constraints, we primarily focused on English-language videos available on the platform. This choice may result in an incomplete understanding of the climate change topic, as users from different cultural backgrounds may express their views on climate change in diverse ways. There is a risk that significant viewpoints and discussions from content in other languages, especially those published in non-English-speaking countries or regions, may be overlooked.

\subsubsection{Insufficient Depth of Comments} The comment analysis in this study was based on qualitative methods, which introduces a degree of subjectivity to the findings. Future research could leverage large language models to analyze the emotional components of comments and derive more objective conclusions. Additionally, the analysis was limited to the \textcolor{blue}{top 6} comments for each video and did not focus on further comments below those comments, which may not fully capture the broader dynamics of the comment sections. Collecting a larger dataset would provide a more comprehensive understanding of audience responses.

\subsubsection{Thematic Limitations} The four characteristics of the video types identified may not be exclusive to the theme of climate change but could also apply to studies of other video subjects. This suggests that the categorisation and analytical framework proposed in the research may require further validation and expansion to accommodate content across different themes. If these characteristics are regarded solely as unique identifiers of climate change discussions, it may lead to the neglect of similar discussion patterns in other themes. This limitation highlights the need to maintain an open perspective when conducting thematic analyses, exploring intersections and differences across various topics to enhance the broad applicability and effectiveness of the research.

\subsubsection{Data Analysis} During the process of data collection and analysis, some collected data may not have been fully utilized. For example, although we retrieved engagement metrics such as the number of views, likes, and shares for each video, we did not conduct an in-depth exploration of these indicators. Similarly, while we defined \textit{Trend Hijacking} videos as those leveraging trending topics for attention, we did not assess whether their use of the climate change tag effectively achieved this purpose. Additionally, the lack of appropriate models limited our ability to apply machine learning techniques to process and interpret video content, which may have resulted in information loss or misinterpretation. Finally, although our sample of 200 videos allowed the construction of a saturated taxonomy, it may not perfectly represent the exact distribution of all video types within the vast \#climatechange ecosystem on TikTok. Future research could employ large-scale computational methods to examine the generalizability and applicability of our categories across diverse cultural, linguistic, and platform contexts.

\subsubsection{Uncertainty Regarding Behavioral Impact} We did not evaluate the effectiveness of the climate change messages in promoting positive behavioral changes. Our research focuses on understanding how creators express their concerns and how interactions unfold in the comment sections, but it does not assess whether these messages lead to actual changes in attitudes or actions. Evaluating the impact of social media messages is crucial for understanding their effectiveness in promoting positive behaviors, such as using big data techniques to track users' subsequent behaviors. This could be a direction for future research, aiming to examine whether creators' messages align with their intended impact.

\section{Conclusion}\label{sec:Conclusion}
Climate research has widely explored how media content, such as scientific data and policy, impacts the audience. However, research on how the public uses video to express personal climate concerns remains limited. This study analyzes \#climatechange videos on TikTok to see how different categories shape the platform's discussion environment. The results highlight four distinct video categories, showcasing the diversity of climate-related content on TikTok. For instance, some users tend to use charts and data to explain climate phenomena or predict their social impact. In contrast, others rely on performances, humor, or activism to convey their values and connect climate change to personal meaning. These creators engage in climate discussions using their preferred expression strategies, enhancing visibility on climate issues in different ways: some try to bridge knowledge gaps for viewers, while others try to foster emotional connections with climate-related topics. Interestingly, audience responses sometimes contradict the context of the videos, such as expressing skepticism, even toward positive calls to action, highlighting the diversity of public values. 

This study reveals the strategies content creators employ to express their concerns about climate change on TikTok, offering valuable insights into how climate change can be discussed on social media. By identifying the trade-off between engagement and scientific accuracy, the research provides a guidance for developing persuasive strategies and structural interventions that align with social media platforms to drive positive climate communication.


\bibliographystyle{ACM-Reference-Format}
\bibliography{CSCW186}

\appendix

\newpage
\section{Appendix}
\label{sec:Appendix}

\label{Appendix}

\begin{figure}[H]
    \centering
    \includegraphics[width=0.6\linewidth]{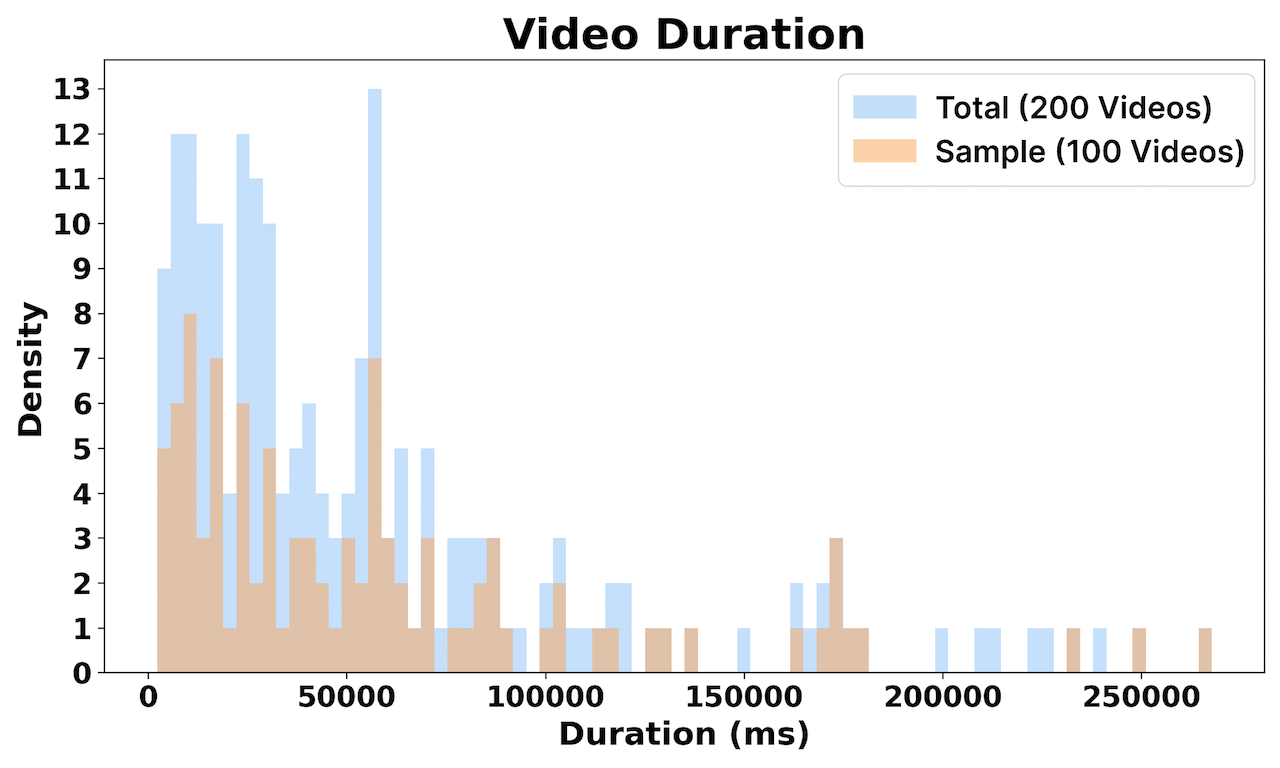}
    \caption{Distribution of duration per video}
    \label{fig66}
\end{figure}

\begin{figure}[H]
    \centering
    \includegraphics[width=0.6\linewidth]{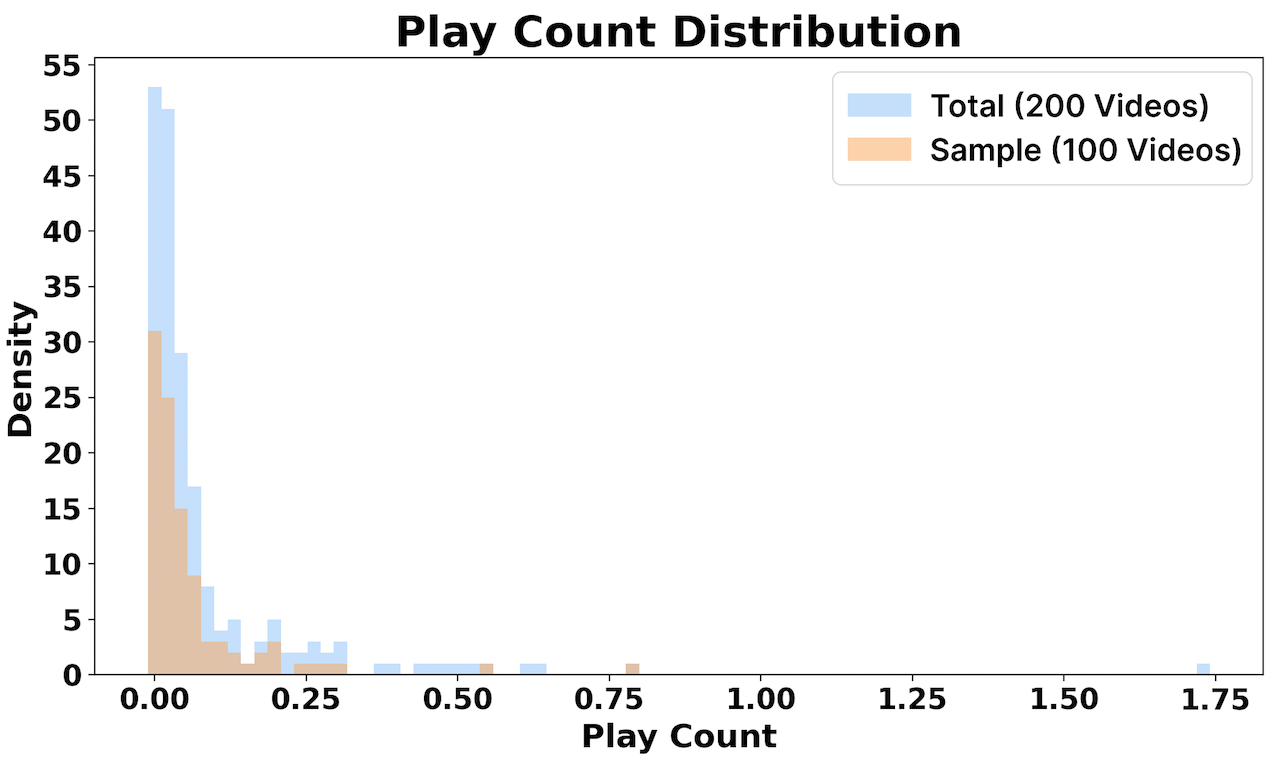}
    \caption{Distribution of play count per video}
    \label{fig67}
\end{figure}

\begin{figure}[H]
    \centering
    \includegraphics[width=0.6\linewidth]{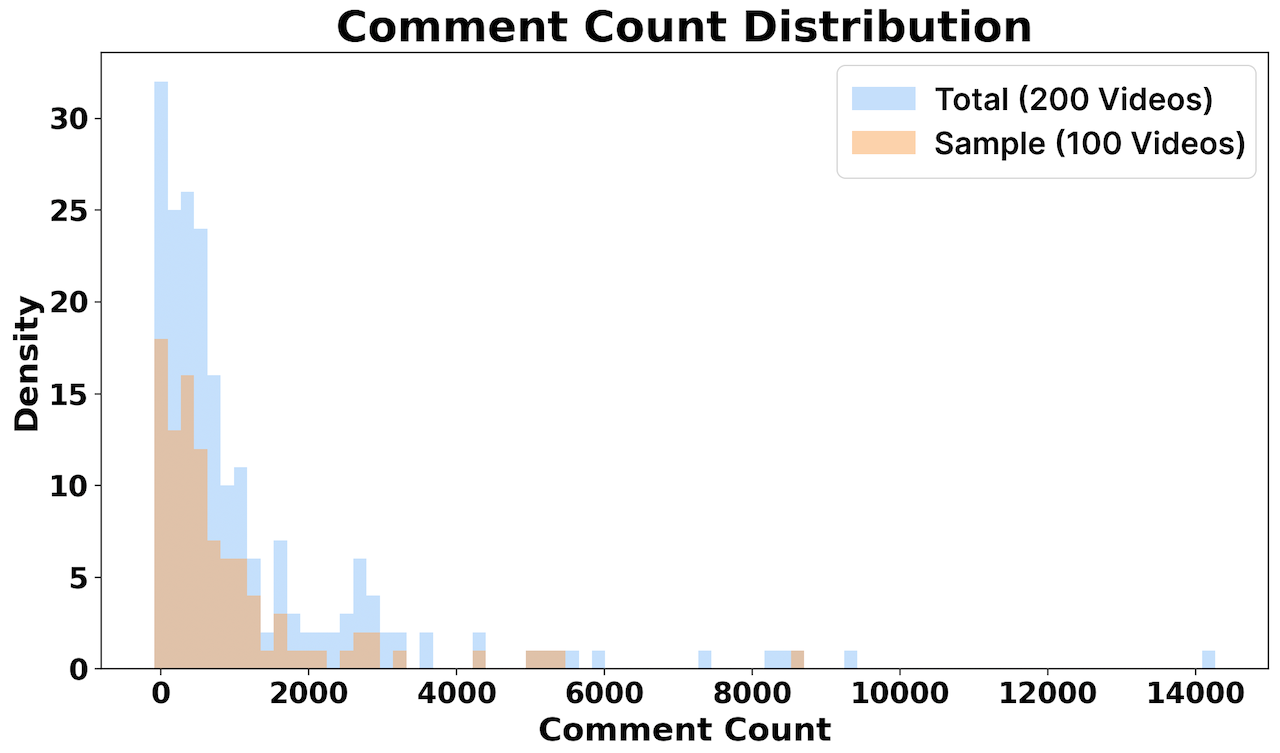}
    \caption{Distribution of comment count per video}
    \label{fig68}
\end{figure}

\begin{figure}[H]
    \centering
    \includegraphics[width=0.6\linewidth]{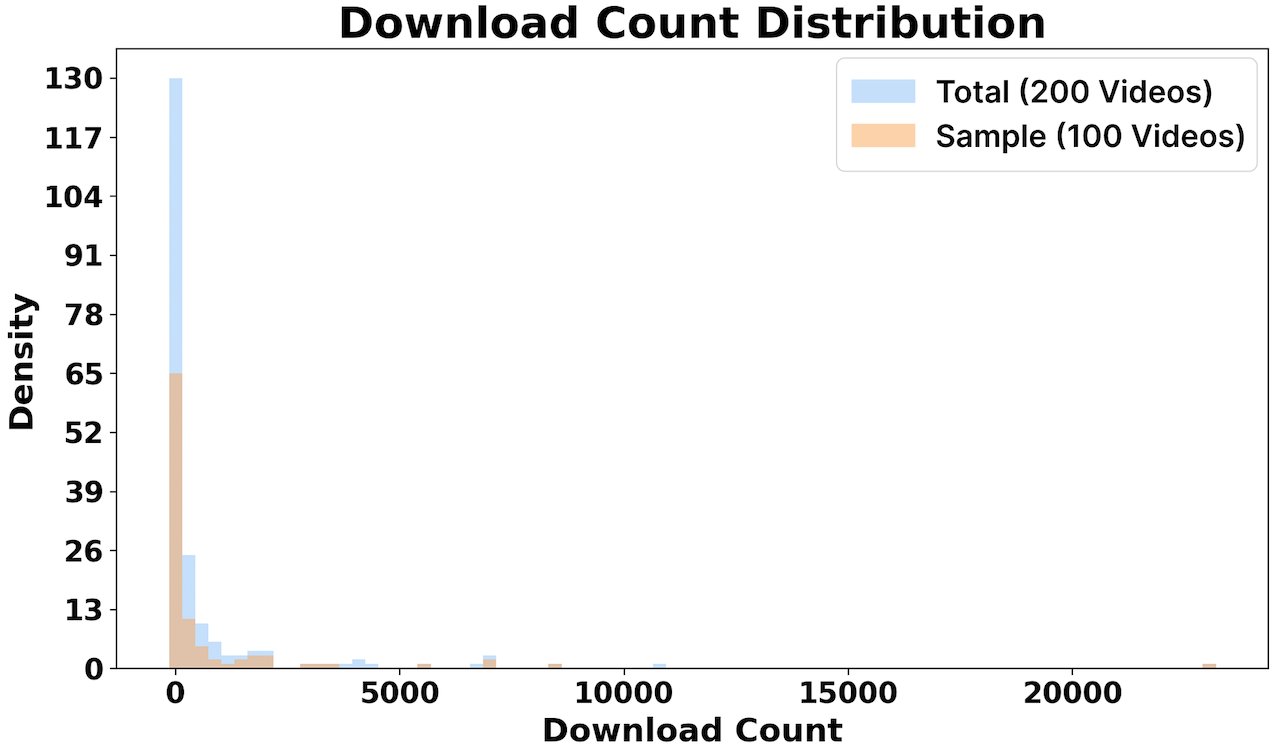}
    \caption{Distribution of download count per video}
    \label{fig69}
\end{figure}

\begin{figure}[H]
    \centering
    \includegraphics[width=0.6\linewidth]{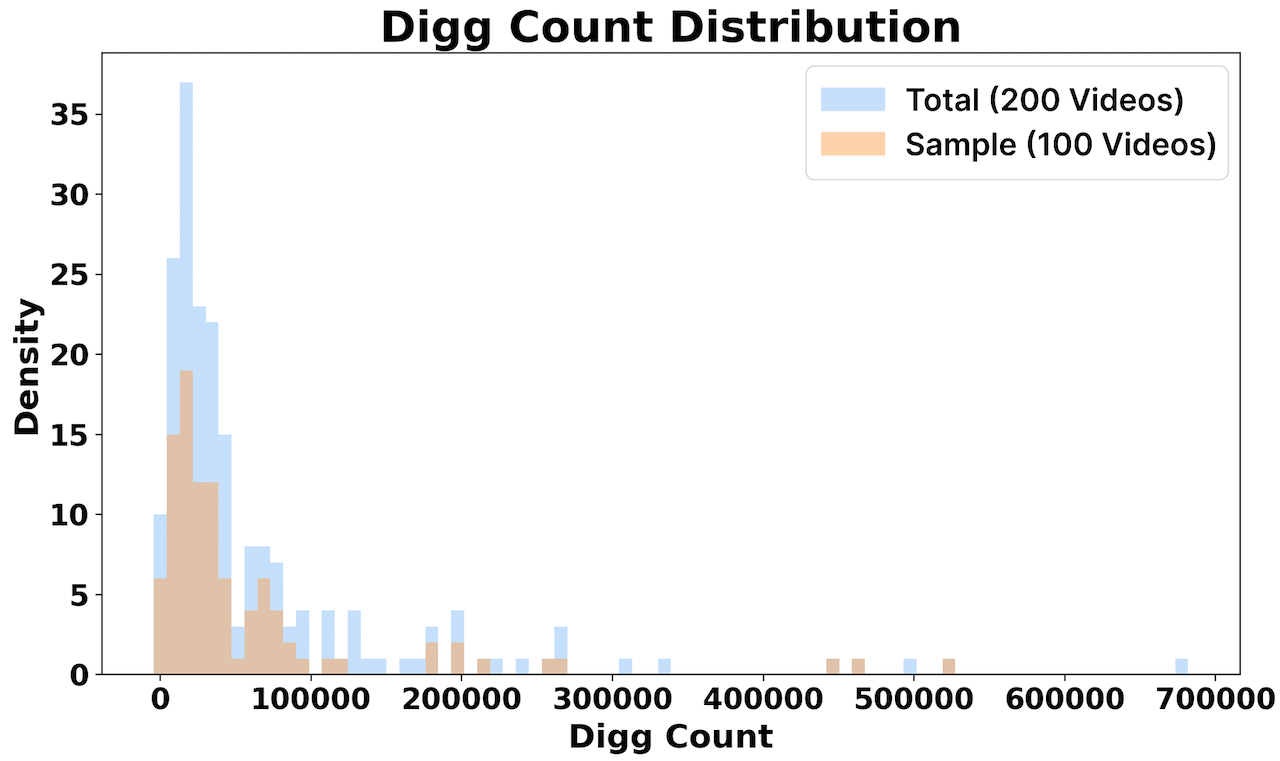}
    \caption{Distribution of digg (like) count per video}
    \label{fig70}
\end{figure}

\begin{figure}[H]
    \centering
    \includegraphics[width=0.6\linewidth]{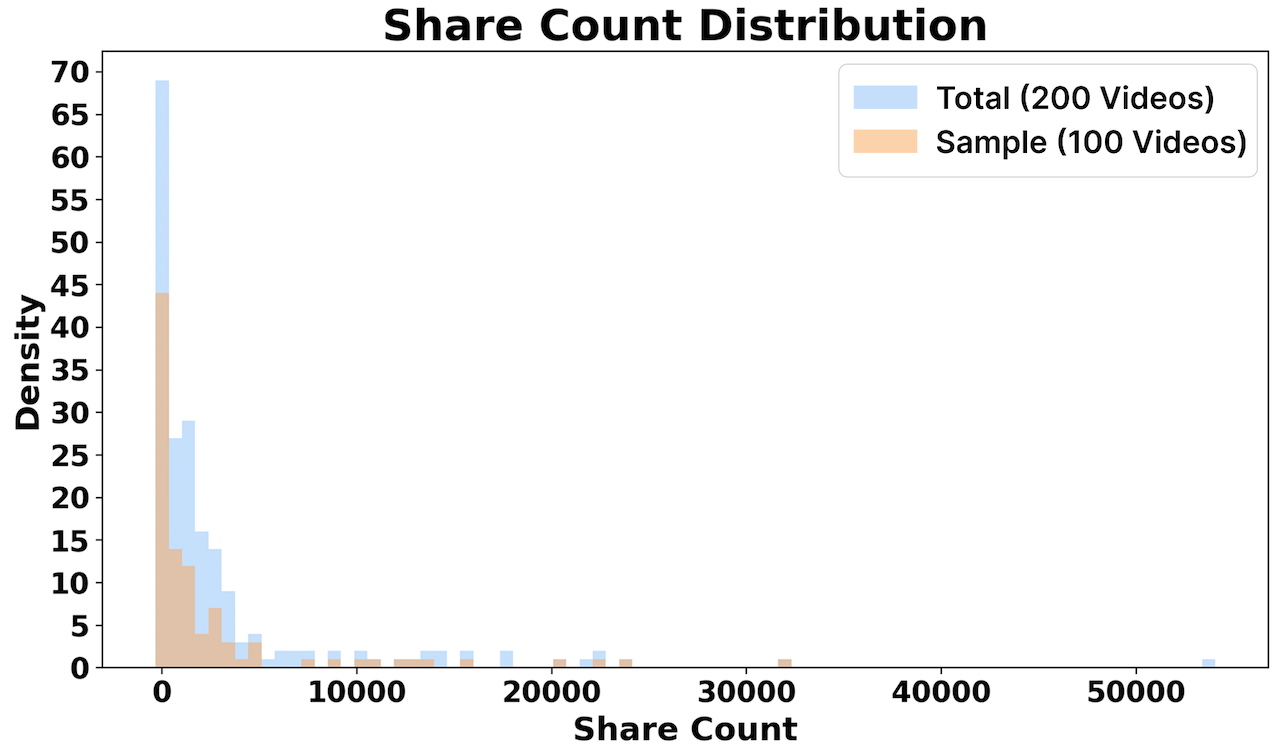}
    \caption{Distribution of share count per video}
    \label{fig71}
\end{figure}

\begin{figure}[H]
    \centering
    \includegraphics[width=0.6\linewidth]{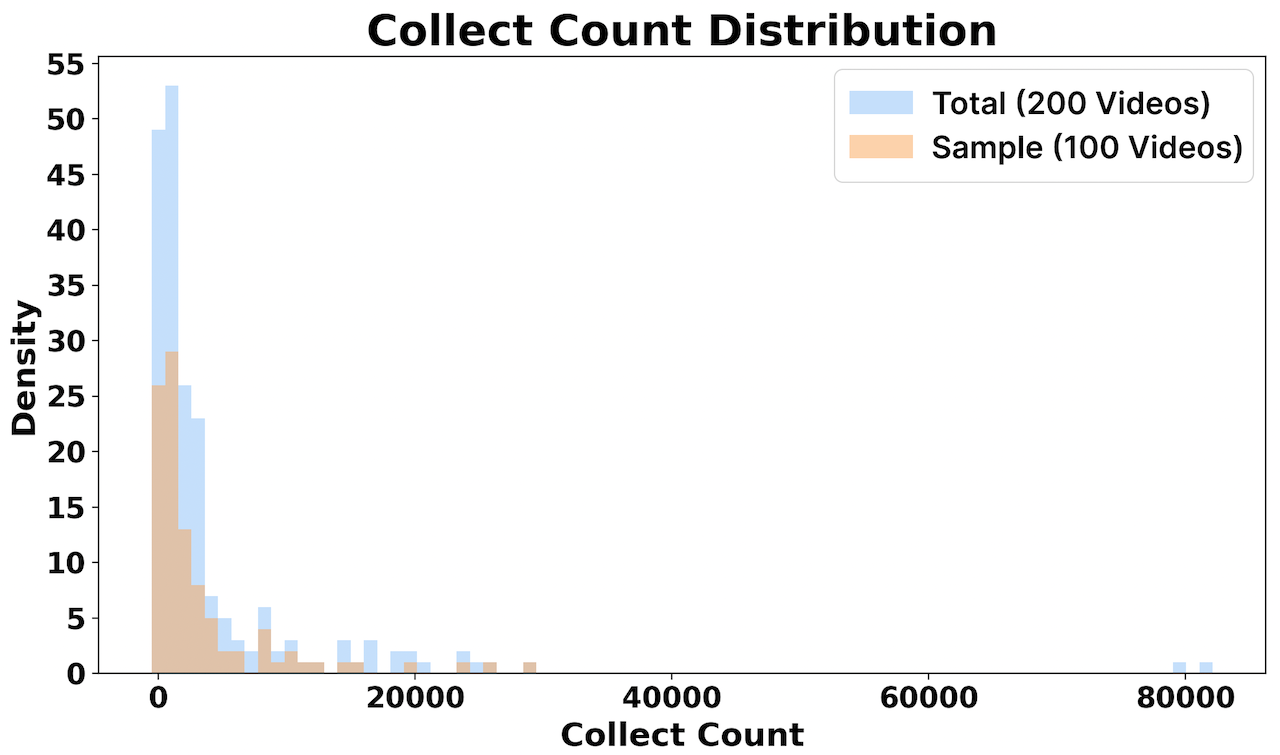}
    \caption{Distribution of collect (save) count per video}
    \label{fig72}
\end{figure}

\end{document}